\documentclass{aastex62}%
\usepackage{amsmath}
\usepackage{amssymb}
\usepackage{amsfonts}%
\usepackage{graphicx}
\begin{document}

\title{Statistical analysis of binary stars from the Gaia catalogue DR2}
\author{Petr Zavada}\email{zavada@fzu.cz}
\author{Karel P\'{\i}\v{s}ka}
\affiliation{Institute of Physics of the Czech Academy of Sciences, Na Slovance 2, 182 21
Prague 8, Czech Republic}
\accepted{by the Astronomical Journal: November 15, 2019}

\begin{abstract}
We have developed a general statistical procedure for analysis of 2D and 3D
finite patterns, which is applied to the data from recently released Gaia-ESA
catalogue DR2. The 2D analysis clearly confirms our former results on the
presence of binaries in the former DR1 catalogue. Our main objective is the
statistical 3D analysis of DR2. For this, it is essential that the DR2
catalogue includes parallaxes and data on the proper motion. The analysis
allows us to determine for each pair of stars the probability that it is the
binary star. This probability is represented by the function $\beta\left(
\Delta\right)  $ depending on the separation. Further, a combined analysis of
the separation with proper motion provides a clear picture of binaries with
two components of the motion: parallel and orbital. The result of this
analysis is an estimate of the average orbital period and mass of the binary
system. The catalogue we have created involves 80560 binary candidates.

\end{abstract}

\affiliation{Institute of Physics of the Czech Academy of Sciences, Na Slovance 2, 182 21
Prague 8, Czech Republic}


\section{Introduction}

In this paper we analyze the recent data from the new catalogue DR2
\cite{gaia2} obtained by the Gaia-ESA mission. If compared with the previous
DR1 catalogue \cite{gaia,gaia1}, the DR2 contains the cleaner data
complemented with parallaxes and data on the 2D proper motion. Parallaxes
allow us to determine the distance of the stars, so we can substantially
enlarge our former DR1 analysis \cite{AApzkp} and work with the 3D patterns of
moving stars.

In the present study, we focus on the statistical analysis of the presence of
binaries. This topic is related to the recent studies of various aspects of
binaries with the use of the catalogues DR1 (\cite{oelkers,semyeong}) and DR2
(\cite{ziegler,esteban}). These authors, apart from their own results, present
an up-to-date overview of important findings on binary stars. Other important
papers exploring wide binaries can be cited from the era before Gaia, for
example \cite{cabalero, close}. However, our approach and objectives are
rather different, so the results obtained are complementary.

Methodology for 2D analysis has been described in detail in our above-quoted
paper. In Sec.\ref{meth} we repeat its essence and perform generalization for
the 3D case. For 2D analysis in Sec.\ref{a2d} we take the same region in DR2
catalogue as we used in the DR1, so we can compare results from both
corresponding data sets.

Principal results are obtained from the 3D analysis of a sample of DR2 data
and are presented in Sec.\ref{ds3d}. This part deals with two issues: the
analysis of 3D separations and the analysis of proper motion of \ pairs of
sources. The combination of both insights provides essential information about
the statistical set of binaries. Obtained results are discussed in
Sec.\ref{disc}. Here we define the probabilistic function $\beta\left(
\Delta\right)  $, which is important for discussion on the occurrence of
binaries. Our present catalogue of binary candidates is described in
Sec.\ref{cat}, where we also shortly discuss its content and overlap with the
catalogue JEC - \cite{esteban}.

The brief summary of the paper is presented in Sec.\ref{sum}. The appendix is
devoted to the derivation of some relations important for our statistical
approach. The most important are distributions of separations of random
sources uniformly distributed inside circles or spheres of unit diameter.
Significant role of these functions for our approach is explained in
Sec.\ref{meth}.

\section{Methodology}

\label{meth}
\begin{figure}[t]
\centering\begin{minipage}{96mm}
\centering
\includegraphics[width=45mm]{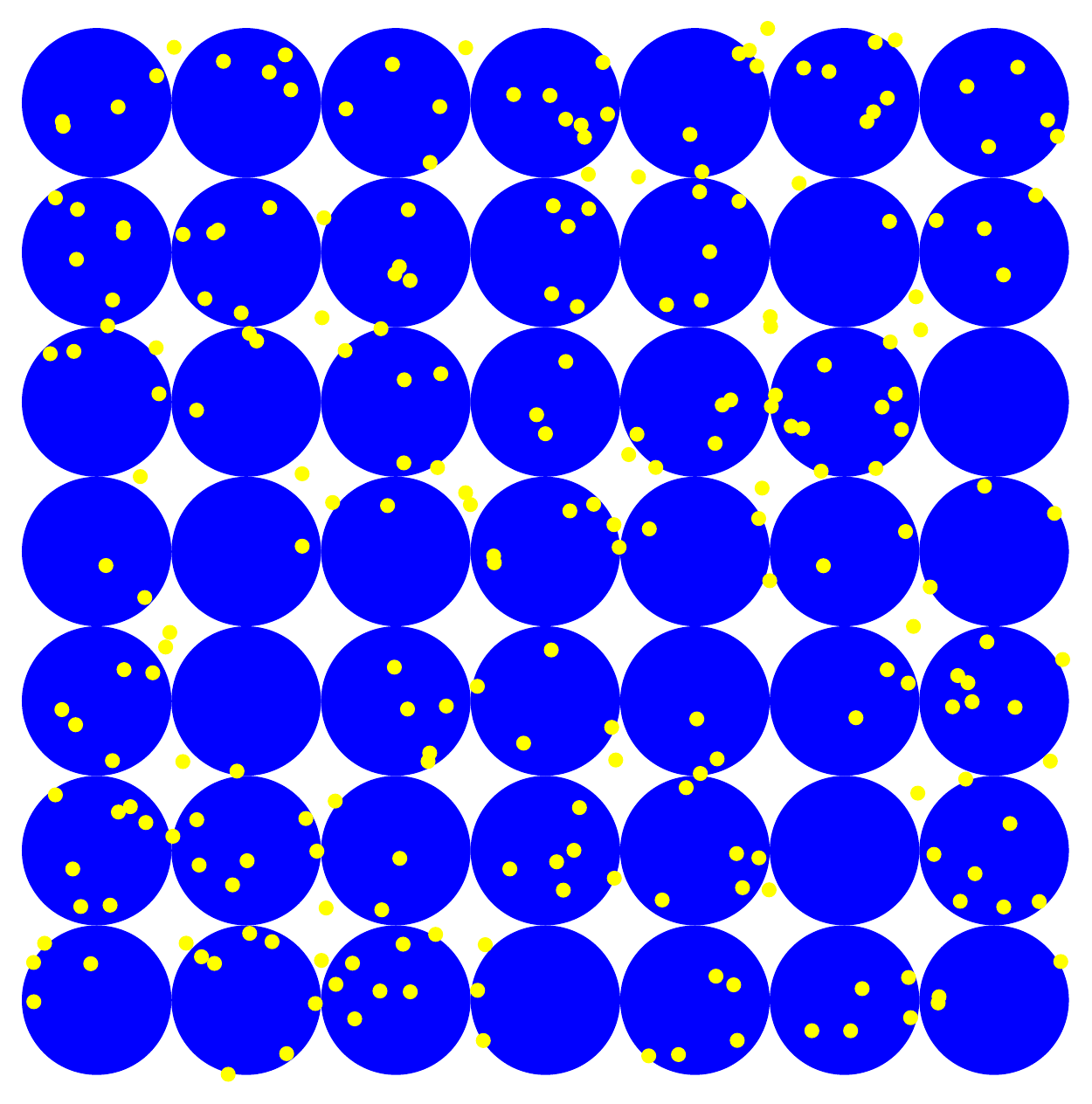}
\includegraphics[width=48mm,clip,trim=0mm 8mm 0mm 4mm]{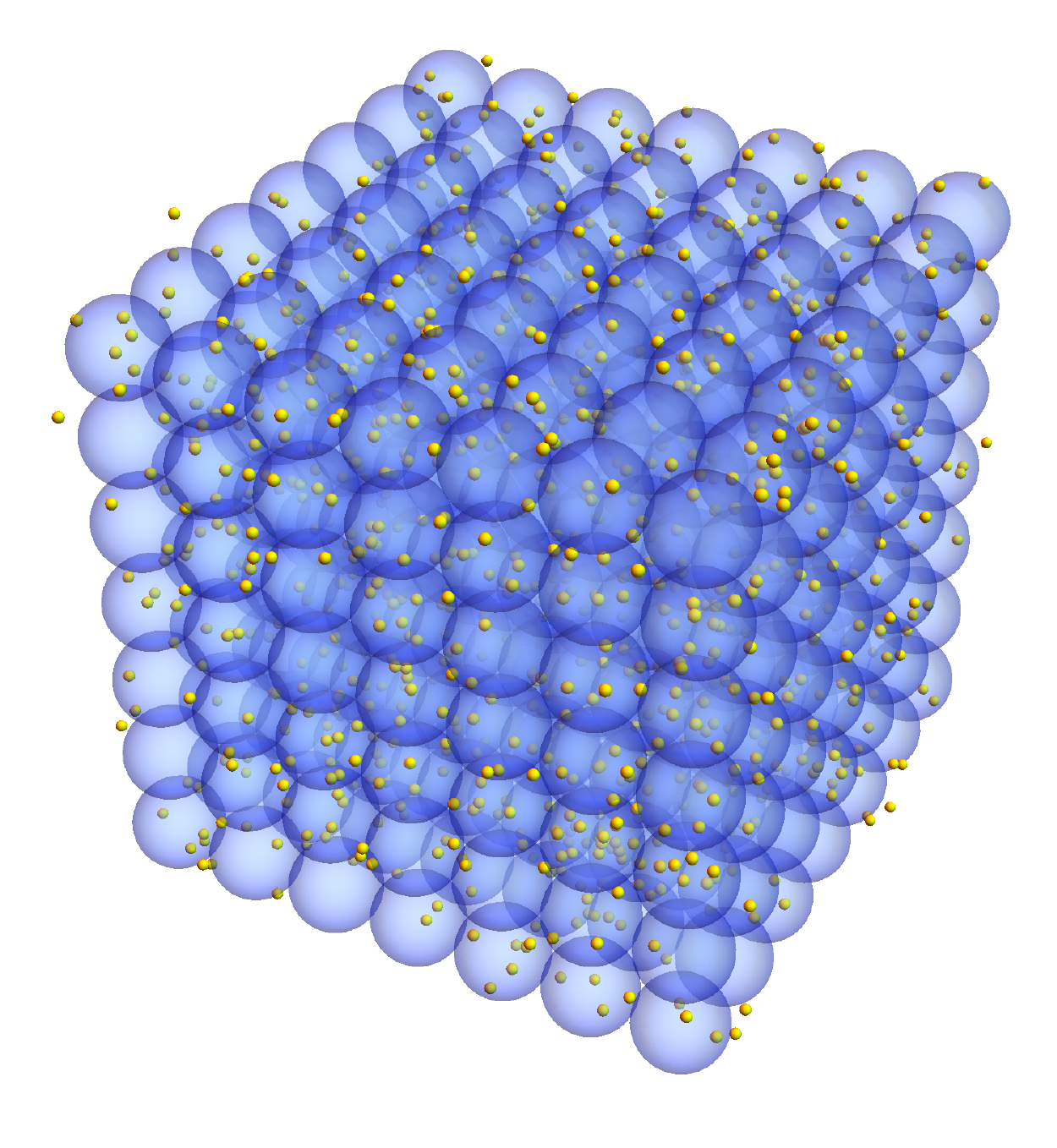}\caption{Grids
of 2D\textit{\ (left)} and 3D \textit{(right)} events with uniform
distributions of the stars.}%
\label{SAB1}%
\end{minipage}\begin{minipage}{44mm}
\centering\includegraphics[width=44mm,trim=37 62 37 0,clip]{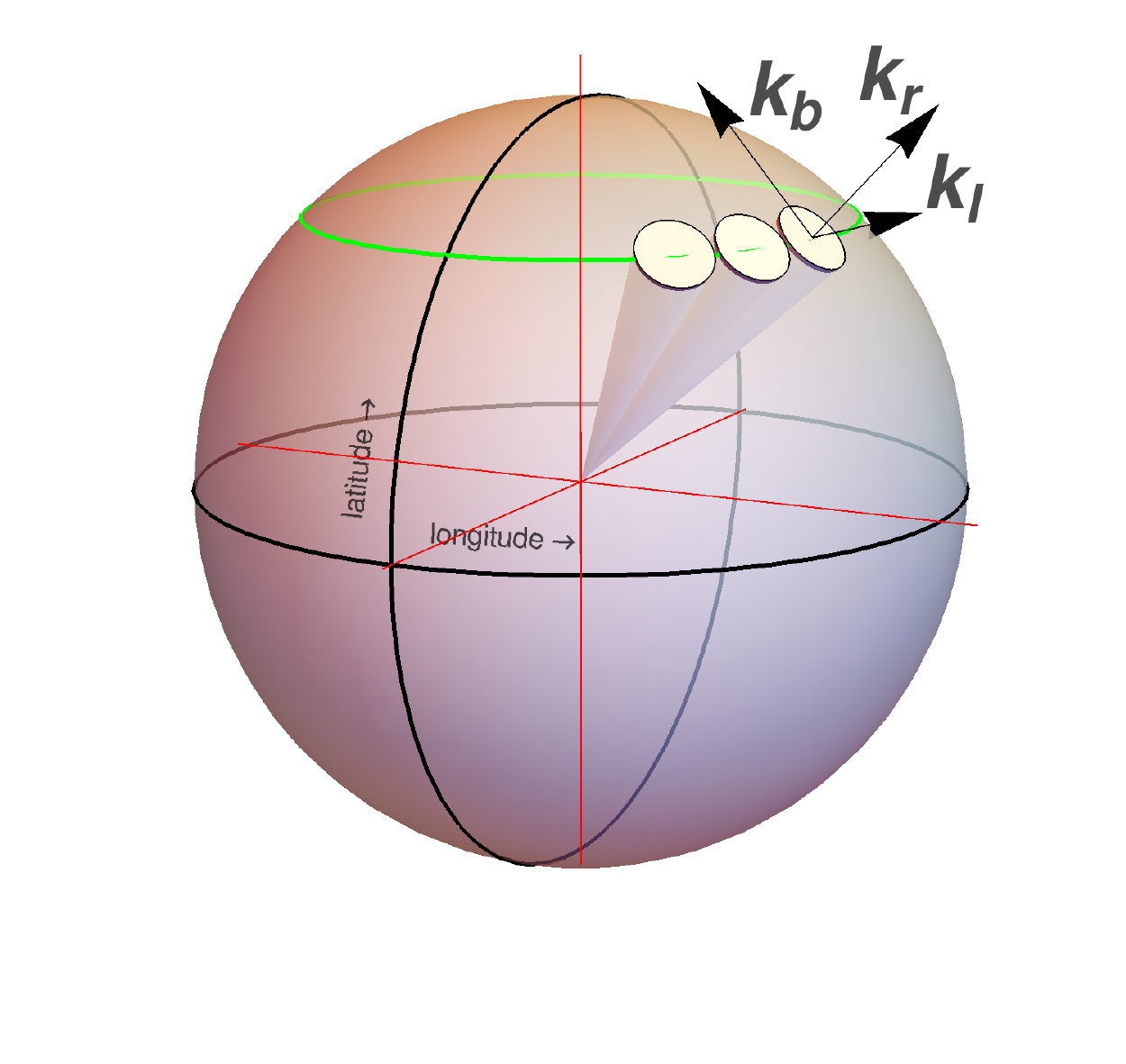}\caption{Galactic reference
frame defined by galactic longitude and latitude with an event local frame defined by
orthonormal basis $\mathbf{k}_{r},\mathbf{k}_{l},\mathbf{k}_{b}$.}%
\label{gala}%
\end{minipage}
\end{figure}

The methods are designed for analysis of the distribution of stars
inside circles or spheres covering the chosen region of the sky, as sketched
in Fig.\ref{SAB1}. These 2D and 3D patterns of stars we call events. Input
data for the generation of the event grids are supposed in the galactic
reference frame. So, the position $\mathbf{L}$ of a source is defined by
spherical coordinates $L,l$ and $b$ (distance from the sun, galactic longitude
and latitude):%
\begin{gather}
\mathbf{L}=L\mathbf{n;\qquad n}=(\cos b\cos l,\cos b\sin l,\sin b),\label{sa1}%
\\
-\frac{\pi}{2}\leq b\leq\frac{\pi}{2},\qquad-\pi<l\leq\pi.\nonumber
\end{gather}
In the centre of circles or spheres we define local orthonormal frame defined
by the basis:
\begin{align}
\mathbf{k}_{r} &  =\mathbf{n}_{0}=(\cos b_{0}\cos l_{0},\cos b_{0}\sin
l_{0},\sin b_{0}),\label{sa3}\\
\mathbf{k}_{l} &  =(-\sin l_{0},\cos l_{0},0),\nonumber\\
\mathbf{k}_{b} &  =(-\sin b_{0}\cos l_{0},-\sin b_{0}\sin l_{0},\cos
b_{0}),\nonumber
\end{align}
where $\mathbf{k}_{r}=\mathbf{n}_{0}\left(  b_{0},l_{0}\right)  $ defines
angular position of the event centre. Unit vector $\mathbf{k}_{l}$ is
perpendicular to $\mathbf{k}_{r}$ and has direction of increasing $l$, see
Fig.\ref{gala}. Unit vector $\mathbf{k}_{b}$ is defined as $\mathbf{k}%
_{b}=\mathbf{k}_{r}\times\mathbf{k}_{l}$ and has direction of increasing $b$.
Vector $\mathbf{k}_{r}$ has radial direction, perpendicular vectors
$\mathbf{k}_{b}$ and $\mathbf{k}_{l}$ \ lies in the transverse plain.

\subsection{Definition of events}

\label{def}The 2D event of the multiplicity $M$ is a set of stars with angular
positions $\mathbf{n}_{i}$\ inside a circle with the event centre
$\mathbf{n}_{0}$ and a small angular radius $\rho_{2}$:
\begin{equation}
\left\vert \mathbf{n}_{i}-\mathbf{n}_{0}\right\vert \leq\rho_{2},\qquad
i=1,...M. \label{sa2}%
\end{equation}
With the use of event local basis (\ref{sa3}), the local coordinates are
defined as%
\begin{equation}
x_{i}=\mathbf{n}_{i}^{\prime}.\mathbf{k}_{l},\qquad y_{i}=\mathbf{n}%
_{i}^{\prime}.\mathbf{k}_{b};\qquad\mathbf{n}_{i}^{\prime}=\mathbf{n}%
_{i}-\mathbf{n}_{0}. \label{sa4}%
\end{equation}
We define 2D event as the set:%
\begin{equation}
\{x_{i},y_{i}\};\quad x_{i}^{2}+y_{i}^{2}\leq\rho_{2}^{2},\quad i=1,...M.
\label{sa5}%
\end{equation}

Since the DR2 catalogue involves also data on parallaxes, we can similarly
generate also the 3D events - patterns of the $M$ sources with position
$\mathbf{L}_{i}$ inside the spheres with the centre $\mathbf{L}_{0}$ and
\ radius $\rho_{3}$:\
\begin{equation}
\left\vert \mathbf{L}_{i}-\mathbf{L}_{0}\right\vert \leq\rho_{3},\qquad
i=1,...M. \label{sa6}%
\end{equation}
With the use of the star positions (\ref{sa1}) and local basis (\ref{sa3}) we
define local coordinates $\{X_{i},Y_{i},Z_{i}\}$ as%

\begin{gather}
X_{i}=\mathbf{N}_{i}.\mathbf{k}_{l},\qquad Y_{i}=\mathbf{N}_{i}.\mathbf{k}%
_{b},\qquad Z_{i}=\mathbf{N}_{i}.\mathbf{k}_{r};\label{SABE4a}\\
\mathbf{N}_{i}=L_{i}\mathbf{n}_{i}-\mathbf{L}_{0},\qquad L_{i}[\mathrm{pc}%
]=\frac{1000}{p_{i}[\mathrm{mas}]}, \label{SABE4b}%
\end{gather}
where $p_{i}$ is parallax, $L_{i}$ is distance\ of the star. 3D event is
defined as the set:%
\begin{equation}
\{X_{i},Y_{i},Z_{i}\};\quad X_{i}^{2}+Y_{i}^{2}+Z_{i}^{2}\leq\rho_{3}%
^{2},\quad i=1,...M. \label{sa8}%
\end{equation}

\subsection{2D methods}

The first method is based on the Fourier analysis of 2D events, where we have
introduced characteristic functions $\Theta_{n}(M)$ depending on the event
multiplicity $M$. These functions are generated by a set of events and measure
statistical deviations from uniform distribution of stars $\left(  \Theta
_{n}(M)=1\right)  $, for instance a tendency to clustering $\left(  \Theta
_{n}(M)>1\right)  $ or anti-clustering $\left(  0<\Theta_{n}(M)<1\right)  $.
Details of the method are described in \cite{AApzkp}, in this paper we will
present only the result.

\begin{figure}[t]
\centering\includegraphics[width=18cm]{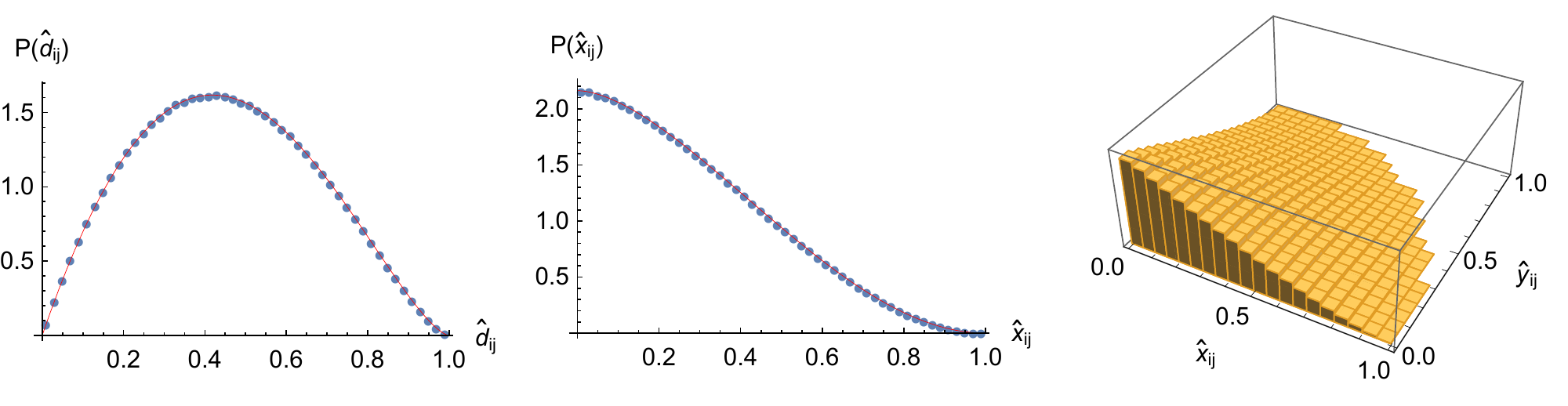}\caption{MC distributions
of separations of uniformly generated stars (points). The red curves represent
the functions (\ref{SABE3}) and (\ref{SABE4}). The Monte-Carlo statistics
corresponds to $5\times10^{5}$ events of multiplicity $M=5.$}%
\label{SAB2}%
\end{figure}

With the use of a second complementary method, we analyze distributions of
angular separations of sources inside the 2D events (\ref{sa5}). Distribution
is generated from the set of events. We use either absolute separations%
\begin{equation}
x_{ij}=\left\vert x_{i}-x_{j}\right\vert ,\quad y_{ij}=\left\vert y_{i}%
-y_{j}\right\vert ,\quad d_{ij}=\sqrt{x_{ij}^{2}+y_{ij}^{2}}, \label{SABE1}%
\end{equation}
or the scaled ones%
\begin{equation}
\hat{x}_{ij}=\frac{x_{ij}}{2\rho_{2}},\quad\hat{y}_{ij}=\frac{y_{ij}}%
{2\rho_{2}},\quad\hat{d}_{ij}=\frac{d_{ij}}{2\rho_{2}}, \label{SABE2}%
\end{equation}
Suitable unit of the parameters $x_{i},y_{i},\rho_{2}$ will be for our purpose
$1as$ ($1^{\prime\prime}$). Distribution of scaled separations generated by
Monte-Carlo (MC) for uniform distribution of stars in the sky is shown in
Fig.\ref{SAB2}. The exact shape of normalized MC distributions reads%
\begin{align}
P(\hat{\xi})  &  =\frac{16\hat{\xi}}{\pi}\left(  \arccos\hat{\xi}-\hat{\xi
}\sqrt{1-\hat{\xi}^{2}}\right)  ;\quad\hat{\xi}=\hat{d}_{ij},\label{SABE3}\\
P(\hat{\xi})  &  =\frac{64}{3\pi^{2}}\left(  (1+\hat{\xi}^{2}%
)\mathrm{EllipticE}(1-\hat{\xi}^{2})-2\hat{\xi^{2}}\mathrm{EllipticK}%
(1-\hat{\xi}^{2})\right)  ;\quad\hat{\xi}=\hat{x}_{ij},\hat{y}_{ij},
\label{SABE4}%
\end{align}
where the functions $\mathrm{EllipticK\ (EllipticE)}$ are complete elliptic
integrals of the first (second) kind. The proof is given in Appendix
\ref{appe}. These distributions do not depend on the event multiplicity and
radius, this is an advantage of the scaled separations. Obviously, we have
always $0<\hat{\xi}<1$. These exact functions replace their approximations
resulting from MC calculation applied in the previous paper.

\subsection{3D methods}

\begin{figure}[t]
\centering\includegraphics[width=18cm]{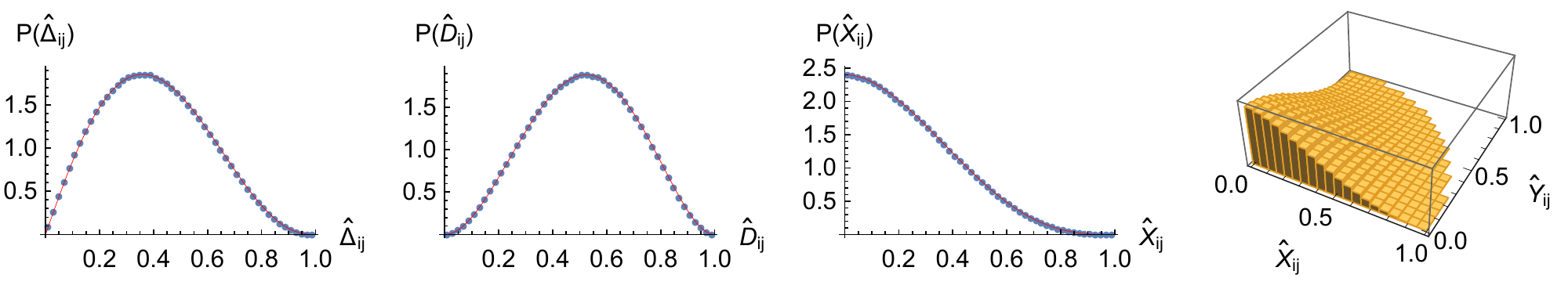}\caption{MC
distributions of separations between uniformly generated stars (points) in 3D.
The red curves represent functions (\ref{SABE7}) - (\ref{SABE8}). The
Monte-Carlo statistics is represented by $5\times10^{5}$ events of
multiplicity $M=5.$}%
\label{SAB3}%
\end{figure}

Similarly, as in the 2D case, we shall work with absolute separations%
\begin{align}
X_{ij}  &  =\left\vert X_{i}-X_{j}\right\vert ,\quad Y_{ij}=\left\vert
Y_{i}-Y_{j}\right\vert ,\quad Z_{ij}=\left\vert Z_{i}-Z_{j}\right\vert
,\label{SABE5}\\
&  D_{ij}=\sqrt{X_{ij}^{2}+Y_{ij}^{2}+Z_{ij}^{2}},\quad\Delta_{ij}%
=\sqrt{X_{ij}^{2}+Y_{ij}^{2}}, \label{SABE5a}%
\end{align}
and/or with the scaled ones%
\begin{equation}
\hat{X}_{ij}=\frac{X_{ij}}{2\rho_{3}},\quad\hat{Y}_{ij}=\frac{Y_{ij}}%
{2\rho_{3}},\quad\hat{Z}_{ij}=\frac{Z_{ij}}{2\rho_{3}},\quad\hat{D}_{ij}%
=\frac{D_{ij}}{2\rho_{3}},\quad\hat{\Delta}_{ij}=\frac{\Delta_{ij}}{2\rho_{3}%
}. \label{SABE6}%
\end{equation}
Suitable unit of the parameters $X_{i},Y_{i},Z_{i},\rho_{3}$ is for our
purpose $1$pc. Distribution of scaled separations generated by MC from the
uniform distribution of stars in 3D region of sky is shown in Fig.\ref{SAB3}.
Exact shapes of these normalized MC distributions read
\begin{align}
P(\hat{\xi})  &  =12\hat{\xi}^{2}(2-3\hat{\xi}+\hat{\xi}^{3});\quad\hat{\xi
}=\hat{D}_{ij},\label{SABE7}\\
P(\hat{\xi})  &  =\frac{9}{2}\hat{\xi}\left(  (2+\hat{\xi}^{2})\sqrt
{1-\hat{\xi}^{2}}+\hat{\xi}^{2}(4-\hat{\xi}^{2})\ln\frac{\hat{\xi}}%
{1+\sqrt{1-\hat{\xi}^{2}}}\right)  ;\quad\hat{\xi}=\hat{\Delta}_{ij}%
,\label{SABE7A}\\
P(\hat{\xi})  &  =\frac{12}{5}(1-\hat{\xi})^{3}(1+3\hat{\xi}+\hat{\xi}%
^{2});\quad\hat{\xi}=\hat{X}_{ij},\hat{Y}_{ij},\hat{Z}_{ij}, \label{SABE8}%
\end{align}
as proved in Appendix \ref{appe}. Shapes of these distributions similarly to
(\ref{SABE3}), (\ref{SABE4}) do not depend on event multiplicity and radius.
The analysis with the use of characteristic functions $\Theta_{n}(M)$ could be
in 3D case done separately in the plains $XY,YZ$ and $ZX$. However, such
analysis is not the aim of the present paper.

\subsection{Aims}

In Sec.\ref{a2d} using the DR2 data set we shall obtain the characteristic
functions $\Theta_{n}(M)$, afterwards we check distributions (\ref{SABE3}) and
(\ref{SABE4}). The distributions (\ref{SABE7})$-$(\ref{SABE8}) will be used
for the data analysis in Sec.\ref{a3d}. All these distributions are of key
importance for the analysis of real data. They represent the templates, which
can reveal a violation of uniformity in the star distributions. Binary (and
multiple) star systems are an example of such a violation, which manifests as
the peaks in the distributions of angular or space separations in the region
of close sources. In general, the scale of expected structure violating
uniformity should be less than the event radius $\rho_{2}$ or $\rho_{3}$.

\section{Analysis of 2D events}

\label{a2d}

\begin{figure}[t]
\centering\includegraphics[width=12cm]{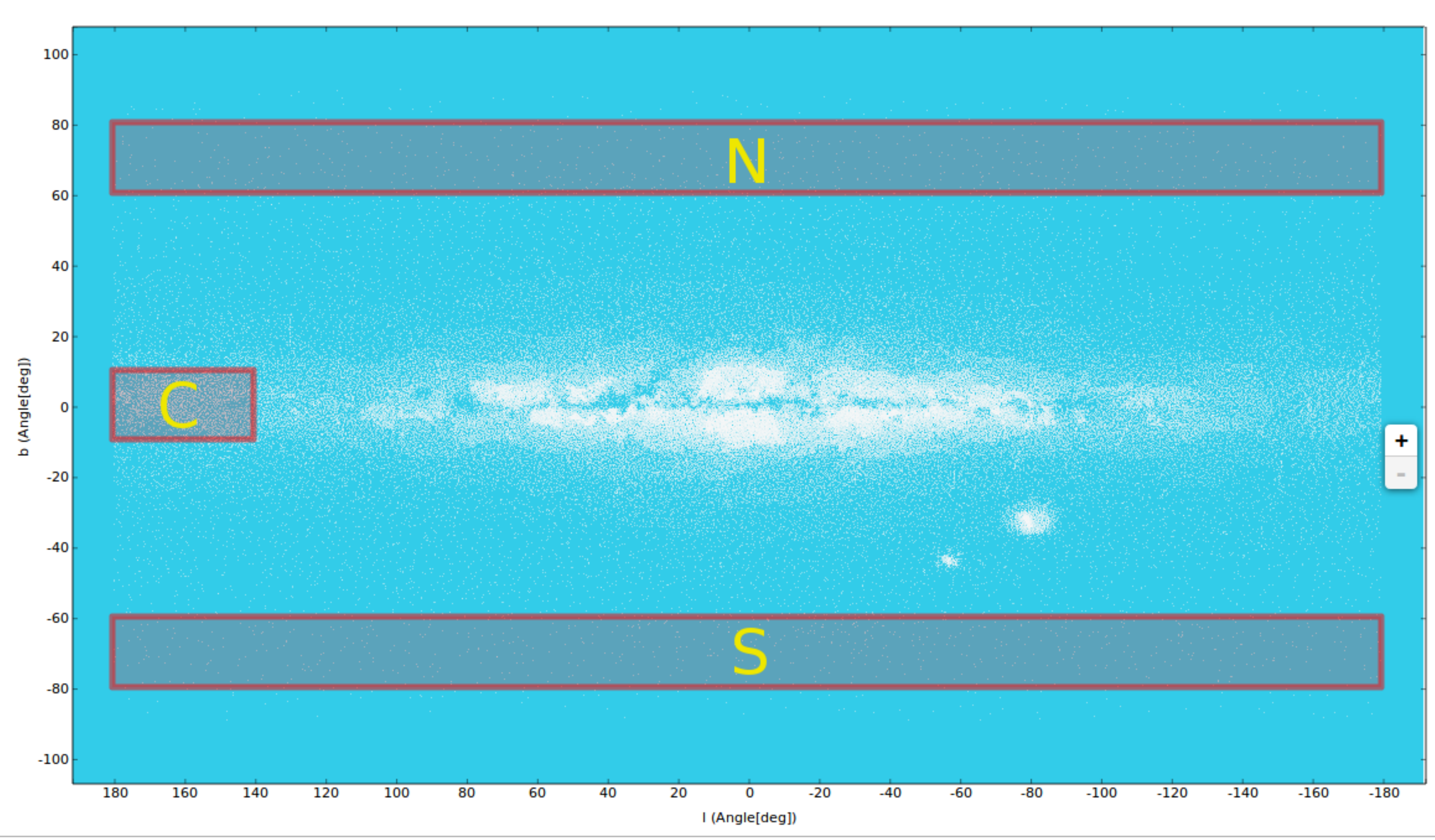}\newline\caption{Analyzed
regions in the DR2 catalogue. }%
\label{faa9}%
\end{figure}

\begin{figure}[t]
\centering\includegraphics[width=18cm]{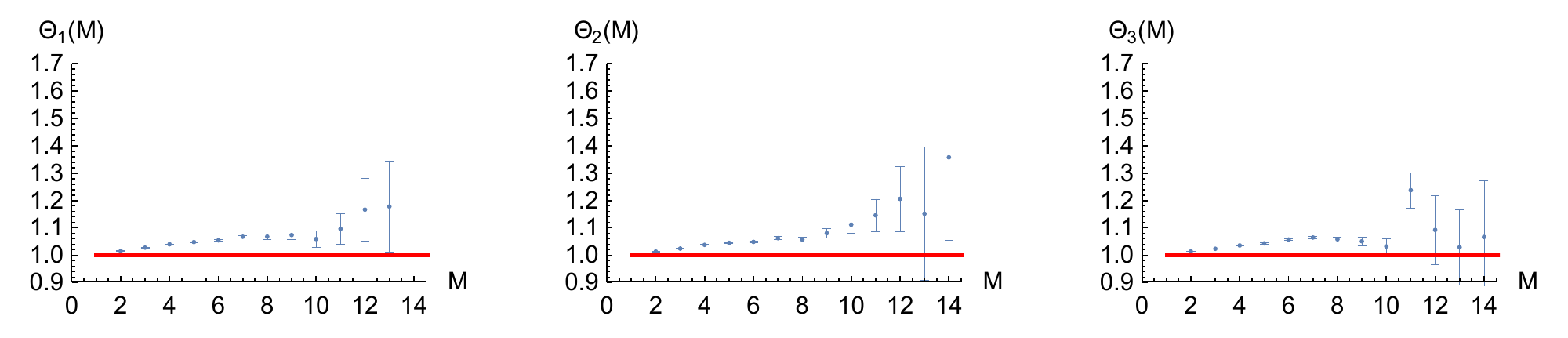}\caption{Characteristic
functions $\Theta_{n}(M)$, $n=1,2,3$\ for events in the area N\&S without any
cut on magnitudes. Red line corresponds to the uniform distribution of
sources.}%
\label{SAB4}%
\end{figure}

Here we present the results obtained from regions of DR2 catalogue listed in
Tab.\ref{tbl1}. The regions are shown in Fig.\ref{faa9}. \begin{table}[ptb]
\begin{center}%
\begin{tabular}
[c]{|c|c|c|c|c|c|}\hline
& 2D region: $l\times b[\deg^{2}]$ & $\rho_{2}[as]$ & $\left\langle
L\right\rangle \left[  \text{pc}\right]  $ & $\left\langle M\right\rangle $ &
$N_{e}$\\\hline
N\&S & $\left\langle -180,180\right\rangle \times\left\langle \pm
60,\pm80\right\rangle $ & $72$ & $1290$ & $3.21$ & $2055674$\\\hline
C & $\left\langle 140,180\right\rangle \times\left\langle -10,10\right\rangle
$ & $18$ & $1912$ & $2.75$ & $3588183$\\\hline
\end{tabular}
\end{center}
\caption{Analyzed regions in the DR2 catalogue, where $\rho_{2}$ is the
angular radius of the events, $\left\langle L\right\rangle ,\left\langle
M\right\rangle $ are average distance and event multiplicity, $N_{e}$ is the
total number of events. Only sources in distance 1-5000pc are taken into
account. The analysis is done \ for events $2\leq M\leq15.$}%
\label{tbl1}%
\end{table}The corresponding events are created with the same angular radius
as in \cite{AApzkp}, which allows us the consistent comparison of results from
the DR1 and DR2 catalogues. First, we checked the events covering the regions
N\&S. Their non-uniformity defined by the characteristic functions $\Theta
_{n}(M)$ is demonstrated in Fig.\ref{SAB4}. The clear result $\Theta_{n}(M)>1$
indicates the presence of clustering.

\begin{figure}[t]
\centering\includegraphics[width=18cm]{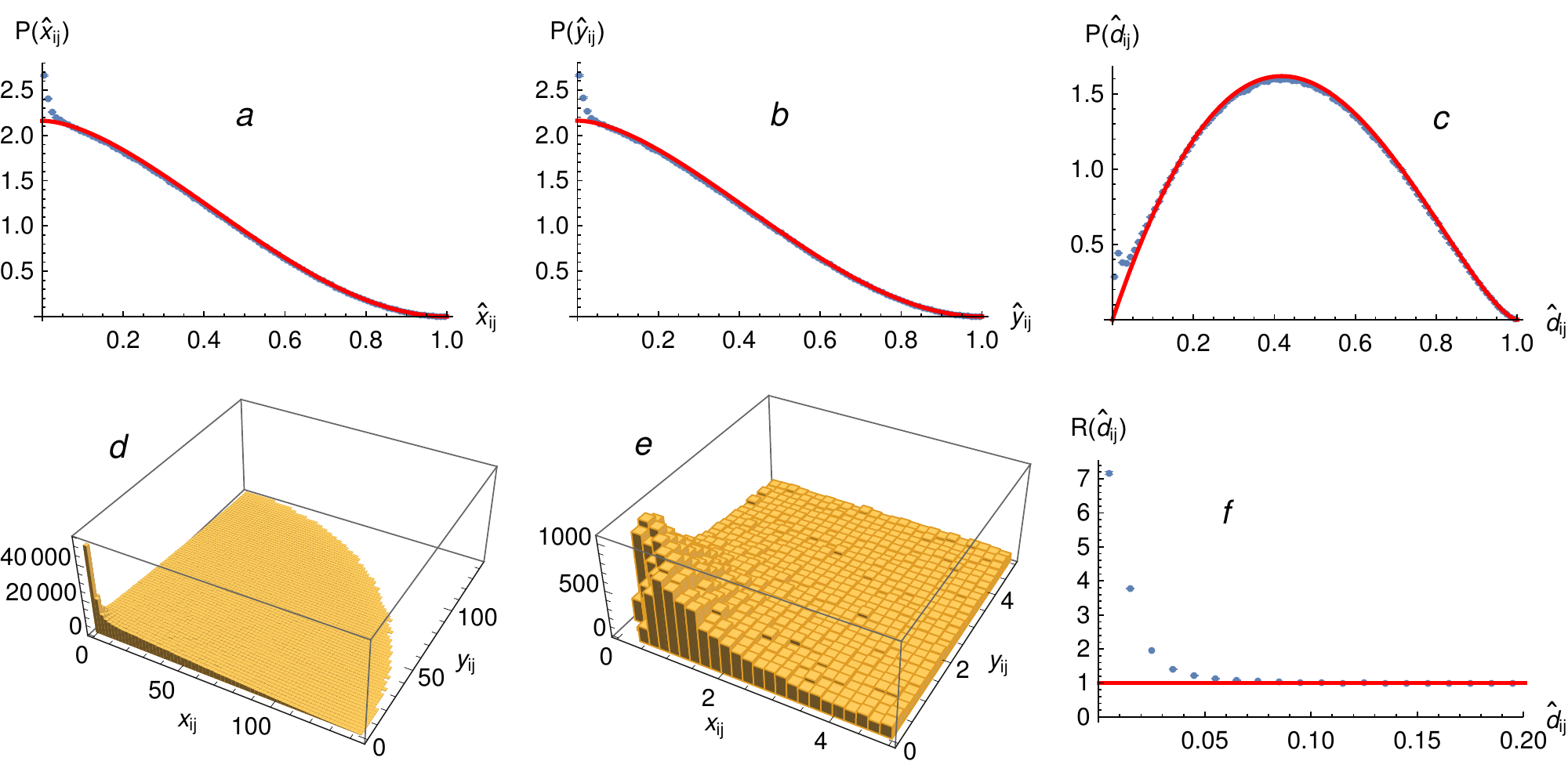}
\includegraphics[width=18cm]{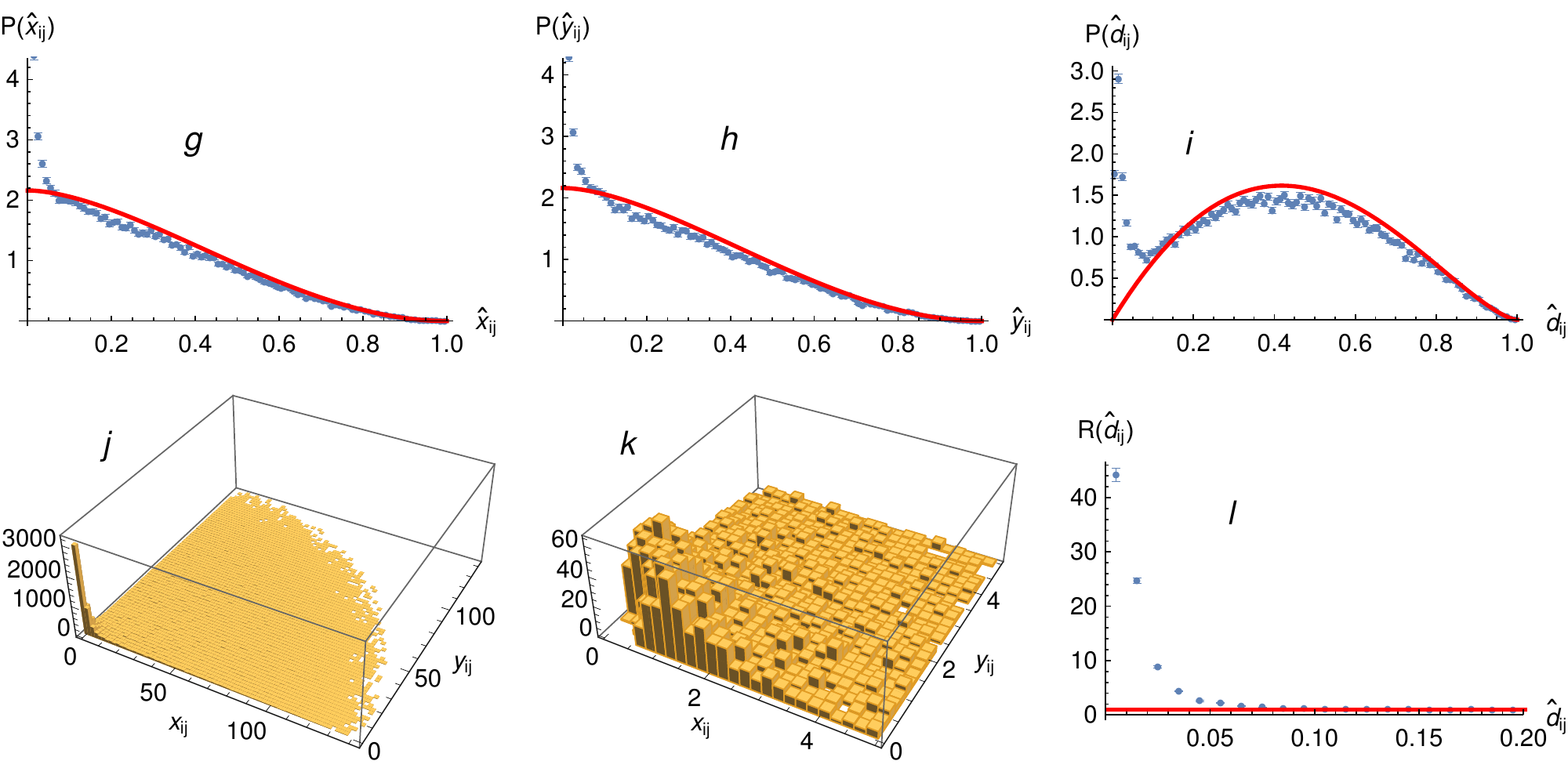}\caption{\textit{Panels a,b,c,d,e,f:
\ }distributions of angular separations in the region N\&S for all $G$. The
blue points in the \textit{panels a,b and c} represent the data on scaled
separations $\hat{x}_{ij},\hat{y}_{ij},\hat{d}_{ij}$ and the red curves are
functions (\ref{SABE3}), (\ref{SABE4}) representing uniform simulation. The
\textit{panel f} is the ratio of data to simulation from \textit{panel c}.
\textit{Panels d and e} represent 3D plot of separations $x_{ij},y_{ij}$ in
different scales (unit$\ $is $1as$). \textit{Panels g,h,i,j,k,l: }the same for
sources $G\leq15.$}%
\label{SAB5}%
\end{figure}

Corresponding distributions of angular separations are shown in
Fig.\ref{SAB5} together with curves (\ref{SABE3}), (\ref{SABE4}).
These results can be compared with those in figures 7 (lower
panels), 10 and 11 in the former paper. We observe:

\begin{figure}[t]
\centering\includegraphics[width=18cm]{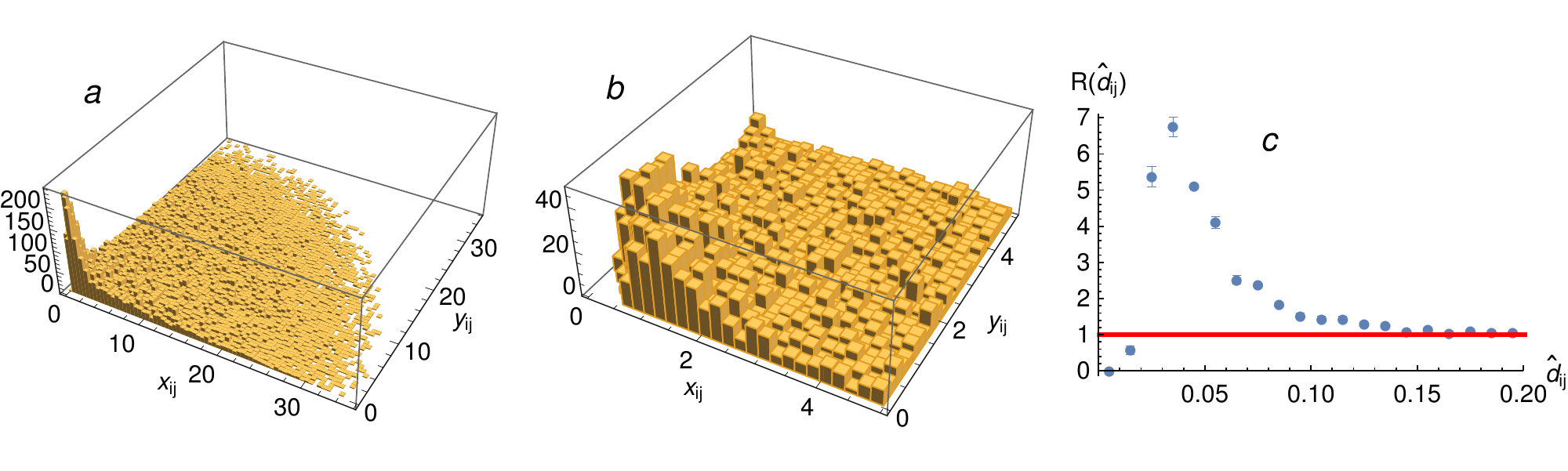}\caption{\textit{Panels
a,b,c: \ }distributions of angular separations in the region C for sources
$G\leq15$. \textit{Panels a }and\textit{\ b} represent 3D plot of separations
$x_{ij},y_{ij}$ in different scales (unit$\ $is $1as$). \textit{Panel c} is
the ratio of data distribution $P(\hat{d}_{ij})$ to the function
(\ref{SABE3}), like \textit{panels f,l} in previous Fig. \ref{SAB5}.}%
\label{SAB6}%
\end{figure}\begin{figure}[h]
\centering\includegraphics[width=17cm]{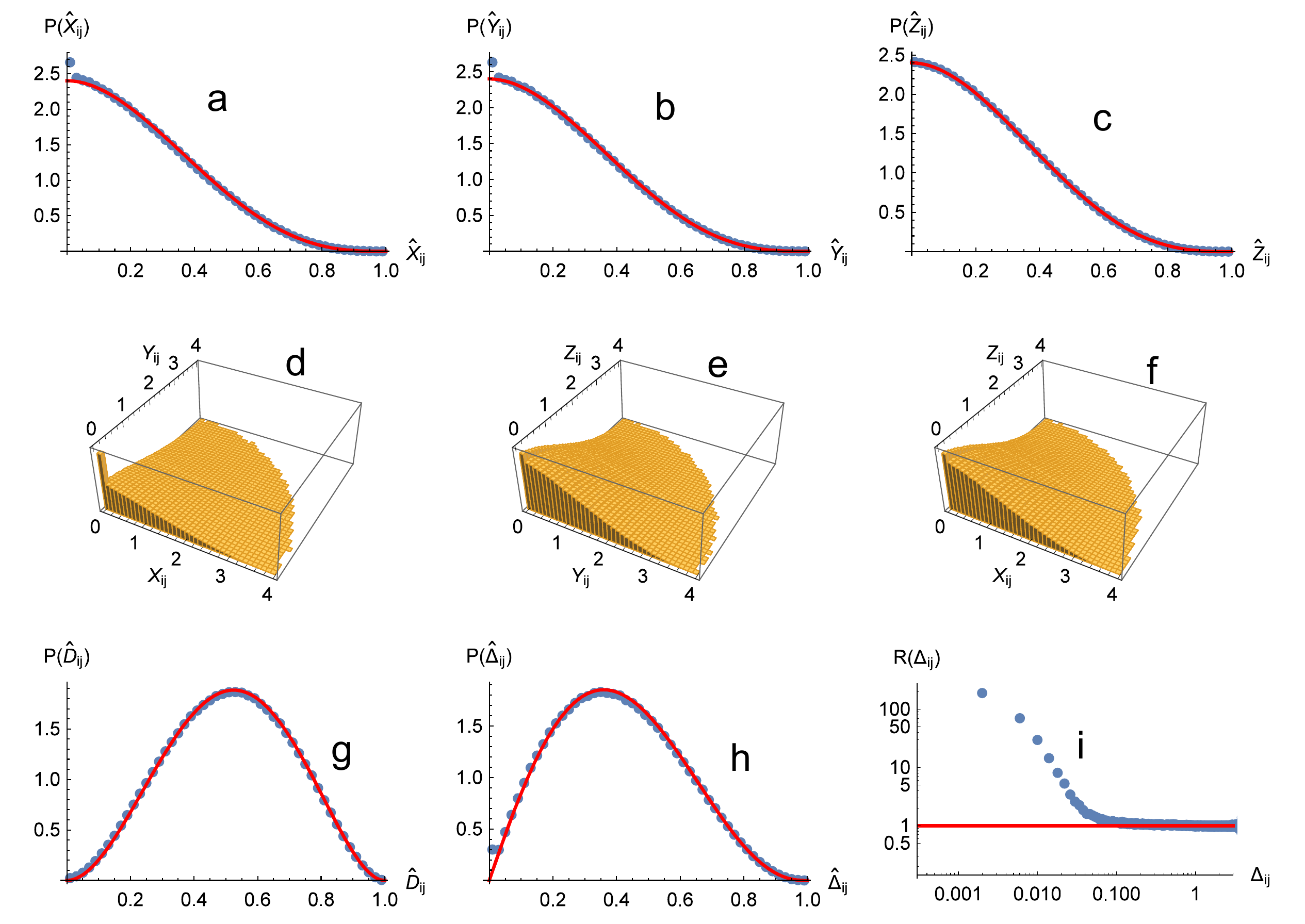}
\includegraphics[width=16cm]{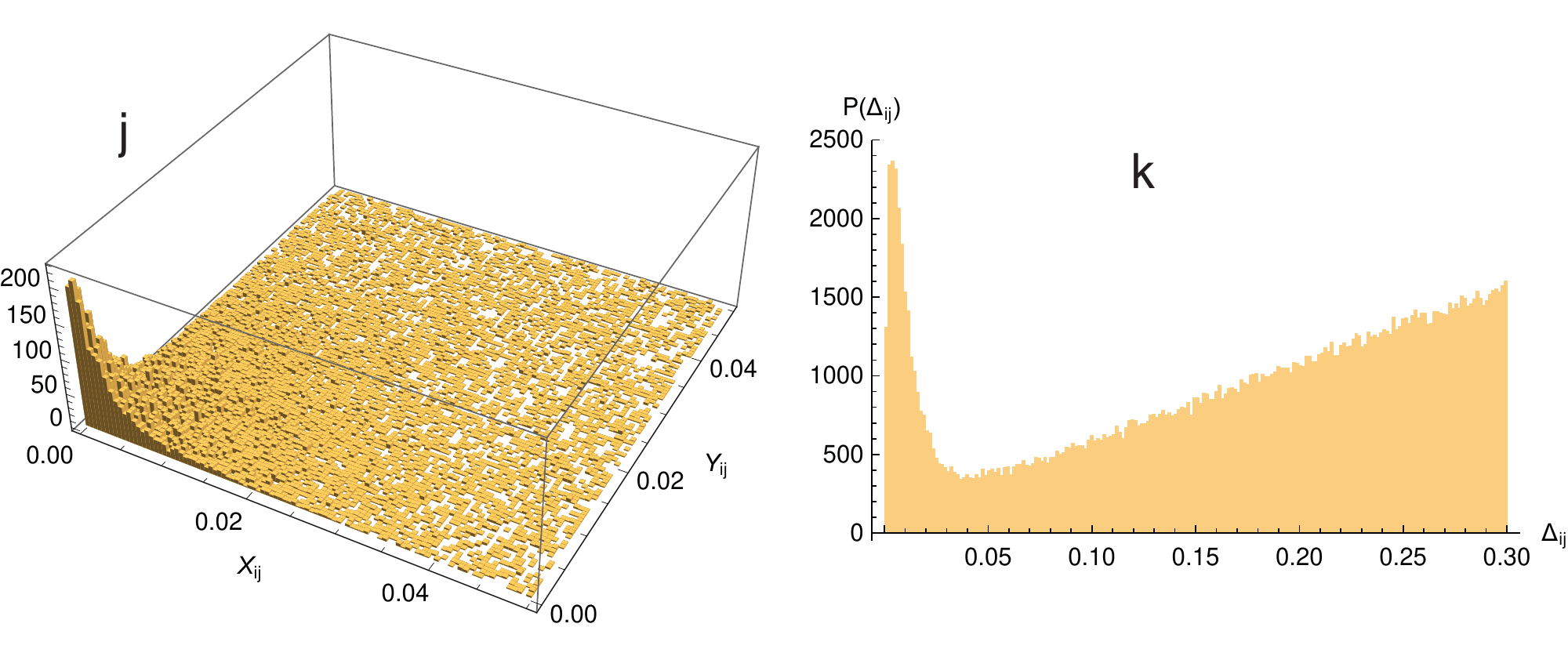}\caption{Distributions of separations
(\ref{SABE5}-\ref{SABE6}) for all $G$. The blue points in the \textit{panels
a,b,c,g and h} represent the data on scaled separations $\hat{X}_{ij},\hat
{Y}_{ij},\hat{Z}_{ij},\hat{D}_{ij},\hat{\Delta}_{ij}$ and the red curves are
functions (\ref{SABE7}-\ref{SABE8}) representing uniform simulation.
\textit{Panels (d,e,f)} show 2D projections of separations. \textit{Panel (i)}
is the ratio of data to simulation from \textit{panel (h)}. \textit{Panels
(j,k)} are magnified version of \textit{(d,h)}. Unit of separations
$X_{ij},Y_{ij},Z_{ij},D_{ij},\Delta_{ij}$ is 1pc. }%
\label{SAB7}%
\end{figure}\begin{figure}[h]
\centering\includegraphics[width=17cm]{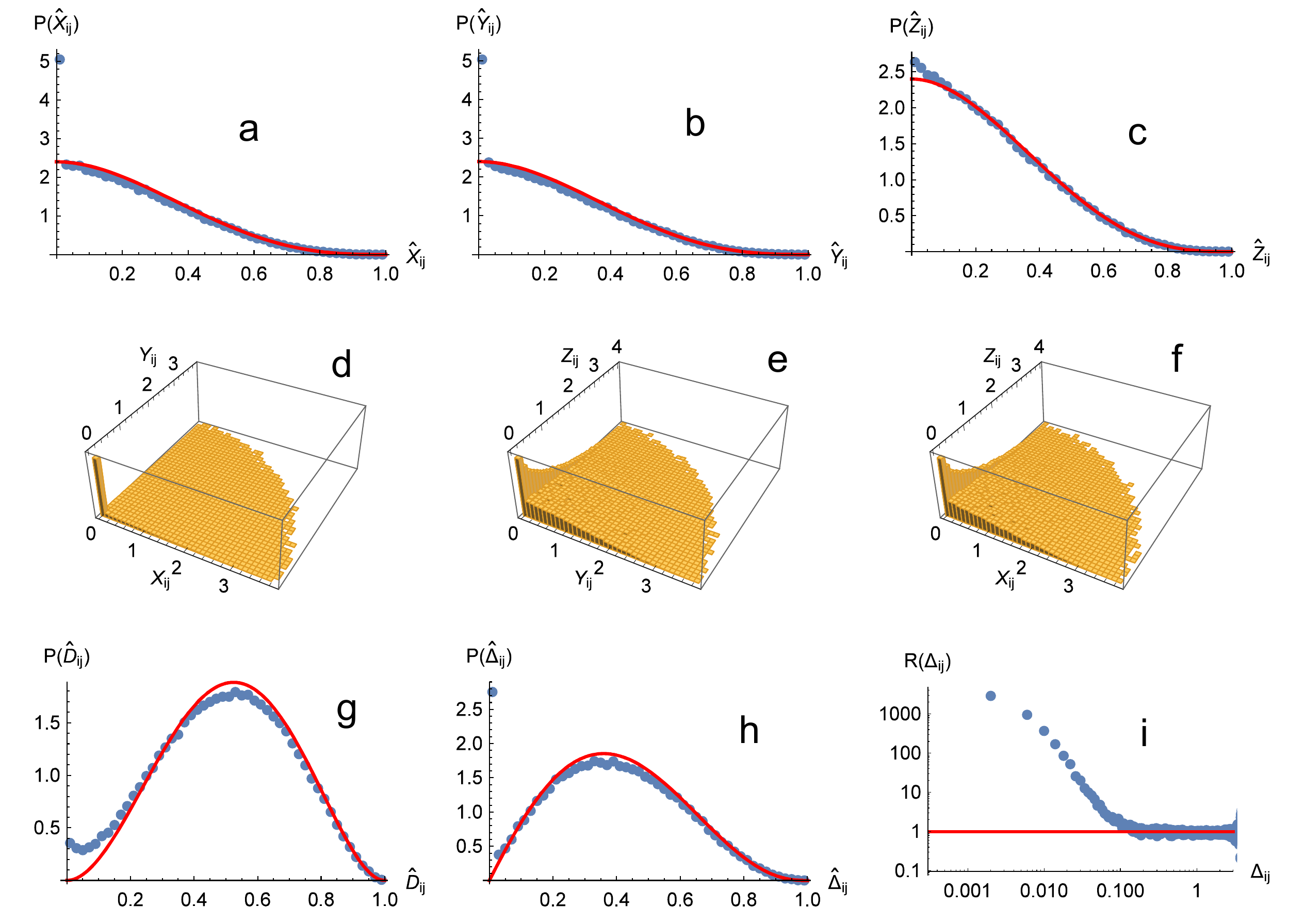}
\includegraphics[width=16cm]{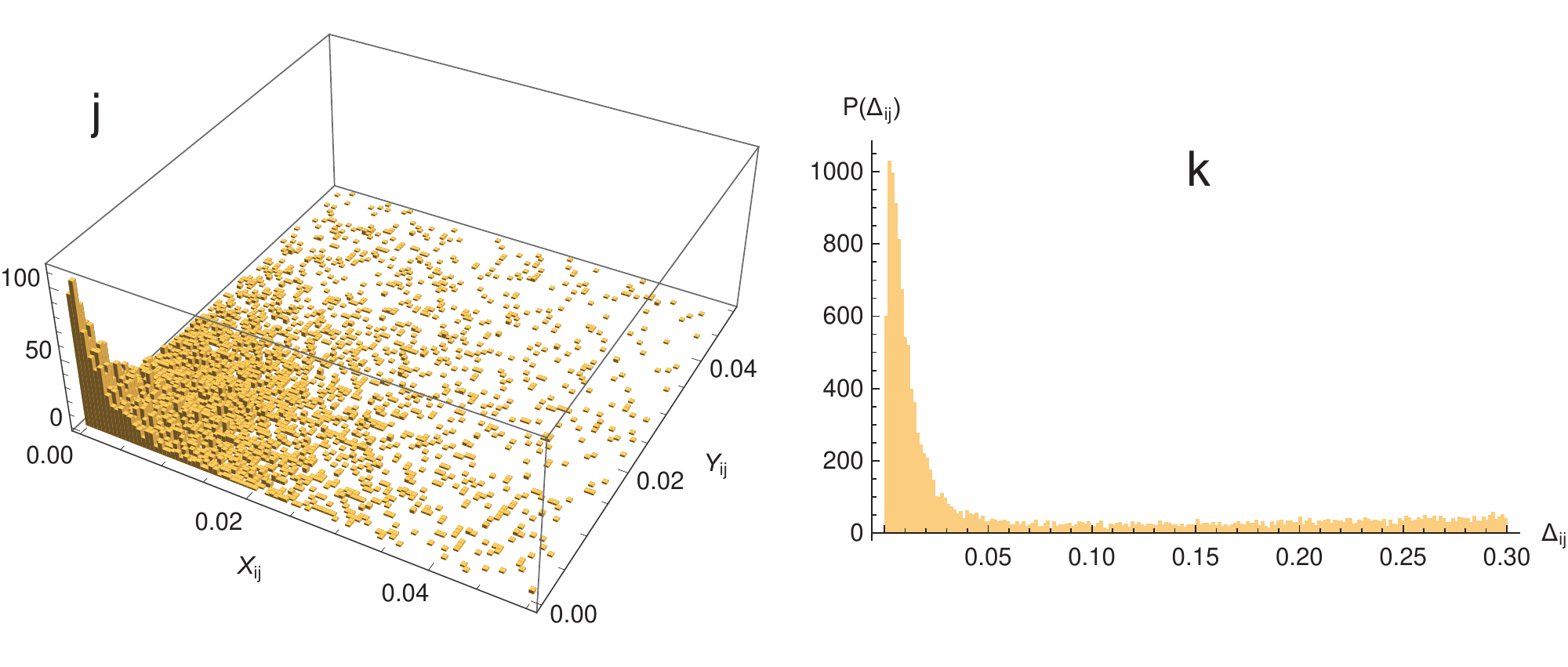}\caption{The same distributions as in
the previous figure, but for bright pairs, $G\leq15.$}%
\label{SAB8}%
\end{figure}

\textit{i)} The peaks at small angular separations in the DR2 corresponding to
binaries are clearer, more pronounced than in the DR1 catalogue. Panels
\textit{e,k} in Fig.\ref{SAB5} demonstrate the double stars separated by
$d_{ij}\lesssim0.5as$ are absent because such close pairs are not resolved in
the DR2 data set as reported in \cite{arenou2}. In both catalogues, we observe
an excess of binaries in the region N\&S for $\hat{d}_{ij}\lesssim0.06$ or
equivalently for $d_{ij}\lesssim8.6as$. For greater separations inside the
event, we observe perfectly uniform distributions of stars. Note the data and
curves are equally normalized for $0<\hat{\xi}<1$. That is why the strong peak
in panels\textit{\ g,h,i} is balanced by a small reduction of distribution
outside the peak. Brighter stars ($G\leq15,$ panels \textit{g,h,i,j,k,l}) show
evidently stronger peaks than sample without any cut on magnitude (panels
\textit{a,b,c,d,e,f}). A similar tendency was observed already in the
catalogue DR1.

\textit{ii)} More pronounced presence of binaries is demonstrated also in
Fig.\ref{SAB4}. The slopes of lines in DR2 are greater than in DR1 -
clustering is more obvious. For $M\leq8$ the slopes are $\approx4\%(8\%)$ for
the DR1(DR2) data. In Fig.\ref{SAB6} we have shown some results obtained in a
more populated region C. Also here we can observe a clear peak at small
angular separations of sources of the magnitude $G\leq15,$ which proves the
presence of binaries. Panel \textit{b} again demonstrates the absence of
double stars separated by $d_{ij}\left(  \hat{d}_{ij}\right)  \lesssim
0.5as\left(  0.014\right)  $ due to insufficient resolution. Different scales
of $\hat{d}_{ij}$ in Figs.\ref{SAB5}\textit{l} and \ref{SAB6}\textit{c} are
due to different radii $\rho_{2}$ of events from N\&S and C regions. Similar
plots could be presented for whole spectrum of magnitudes in the region C,
however elevation above the red line due to binaries is much less than that in
Fig.\ref{SAB6}c. The reason can be that denser region C with all magnitudes
generates higher background of the optical doubles and a consequently lower
relative rate of binaries.

\section{Binaries in 3D events}

\label{ds3d}

We present the results obtained from 3D region defined in Tab.\ref{tab1}.
\begin{table}[ptb]
\begin{center}%
\begin{tabular}
[c]{|c|c|c|c|c|}\hline
Region & $\rho_{3}\left[  \text{pc}\right]  $ & $\left\langle L\right\rangle
\left[  \text{pc}\right]  $ & $\left\langle M\right\rangle $ & $N_{e}$\\\hline
cube of edge 400pc & 2 & 188 & 6.8 & 727744\\\hline
\end{tabular}
\end{center}
\caption{Region of 3D analysis is cube centred at the origin of the galactic
reference frame. Only sources of positive parallax are included. $\rho_{3}$ is
the radius of events, $\left\langle L\right\rangle ,\left\langle
M\right\rangle $ are average distance and event multiplicity, $N_{e}$ is the
total number of events. The analysis is done for events $2\leq M\leq15.$}%
\label{tab1}%
\end{table}The parallax and angular components of the star proper motion are
the parameters, which substantially enrich the recent Gaia data. We work with
the 3D events (\ref{sa8}).

\subsection{Analysis of 3D separations}

\label{a3d}

The summary results of the analysis obtained from all magnitudes $G$ are shown
in Fig.\ref{SAB7}. Distributions of scaled separations in panels
\textit{a,b,c,g,h} perfectly match the uniform distribution of sources, but
with exception of the first bin in \textit{a,b,h.} Apparent excess of very
close pairs in planes $XY,YZ,ZX$ is seen also in panels\textit{\ d,e,f.
}\ However the sharp peak can be observed only in the plane $XY$ (panel
\textit{d }and its magnified version \textit{j}). Smearing in direction of $Z$
(difference of radial positions, panels\textit{\ c,e,f}) is due to lower
accuracy in measuring of parallaxes. The errors of local coordinates depend on
the errors of separations and with the use of definitions (\ref{SABE4a}) and
(\ref{SABE4b}) are calculated as%
\begin{align}
\delta X_{ij}  &  =\sqrt{\left(  \frac{\partial X_{ij}}{\partial L_{i}}\delta
L_{i}\right)  ^{2}+\left(  \frac{\partial X_{ij}}{\partial L_{j}}\delta
L_{j}\right)  ^{2}}\label{SABE8a}\\
&  \approx\delta L\sqrt{\left(  \mathbf{n}_{i}\mathbf{k}_{l}\right)
^{2}+\left(  \mathbf{n}_{j}\mathbf{k}_{l}\right)  ^{2}}\leq\sqrt{2}\rho
_{3}\frac{\delta L}{L}\nonumber
\end{align}
and similarly%
\begin{equation}
\delta Y_{ij}\leq\sqrt{2}\rho_{3}\frac{\delta L}{L},\qquad\delta Z_{ij}%
\approx\delta L\sqrt{\left(  \mathbf{n}_{i}\mathbf{k}_{r}\right)  ^{2}+\left(
\mathbf{n}_{j}\mathbf{k}_{r}\right)  ^{2}}\leq\sqrt{2}\delta L. \label{SABE8b}%
\end{equation}
Note that $\delta L/L=\delta p/p$, where $p$ is parallax. Obviously $\delta
Z_{ij}\gg$ $\delta X_{ij},\delta Y_{ij}$ since $\mathbf{n}_{i}\mathbf{k}%
_{l},\mathbf{n}_{i}\mathbf{k}_{b}\ll\mathbf{n}_{i}\mathbf{k}_{r}\lessapprox1$.
That is why we prefer distributions of $\Delta_{ij}$ to $D_{ij}$ for obtaining
precise results. Maximum values of separations $X,Y,Z$ is 4pc, which follows
from the event radius $\rho_{3}=$2pc. The excess of close pairs is obvious
most explicitly from the distribution of $\Delta_{ij}$ in panel\textit{\ k
}that represents a magnified version of \textit{h}. The important ratio of the
distribution $P(\Delta_{ij})$ to the uniform simulation (red curve in
panel\textit{\ h }rescaled to $\Delta_{ij}=2\rho_{3}\hat{\Delta}_{ij}$)
\textit{\ }is shown in logarithmic scale of $\Delta_{ij}$ in panel \textit{i.}

The same distributions but for brighter sources $G\leq15$ are shown in
Fig.\ref{SAB8}. Similarly, as in the 2D case, the peaks are stronger for
brighter sources and distributions outside the peaks confirm uniformity of the
star distribution. Again due to equal normalization of data and red curves for
$0<\hat{\xi}<1$ the strong peak in panels\textit{\ g,h} is balanced by a small
reduction of distribution beyond the peak. The excess of close pairs observed
in both figures again convincingly indicates the presence of binaries.

For quantitative estimates, the important panels are \textit{i}, which display
the ratio data/simulation. This is more accurate than only displaying peaks
with some undefined background. Panels\textit{\ i} suggest that separations of
binary systems in the analyzed region meet very approximately
\begin{equation}
\ \Delta_{ij}\lesssim0.1-0.2\text{pc}. \label{SABE9}%
\end{equation}
We observe only a tail of distribution corresponding to more separated
binaries. Closer pairs are absent due to the limited angular resolution in DR2
data. This result is compatible with the older data reported in \cite{close}.
In Sec.\ref{disc} a more discussion is devoted to the probability of the
binary separation above this limit. We have checked that sampling with events
generated by spheres of different radius ($\rho_{3}=5$pc) does not change the
approximate result (\ref{SABE9}).

\subsection{Proper motion of binaries}

\label{pmds}
\begin{figure}[t]
\centering\includegraphics[width=18cm]{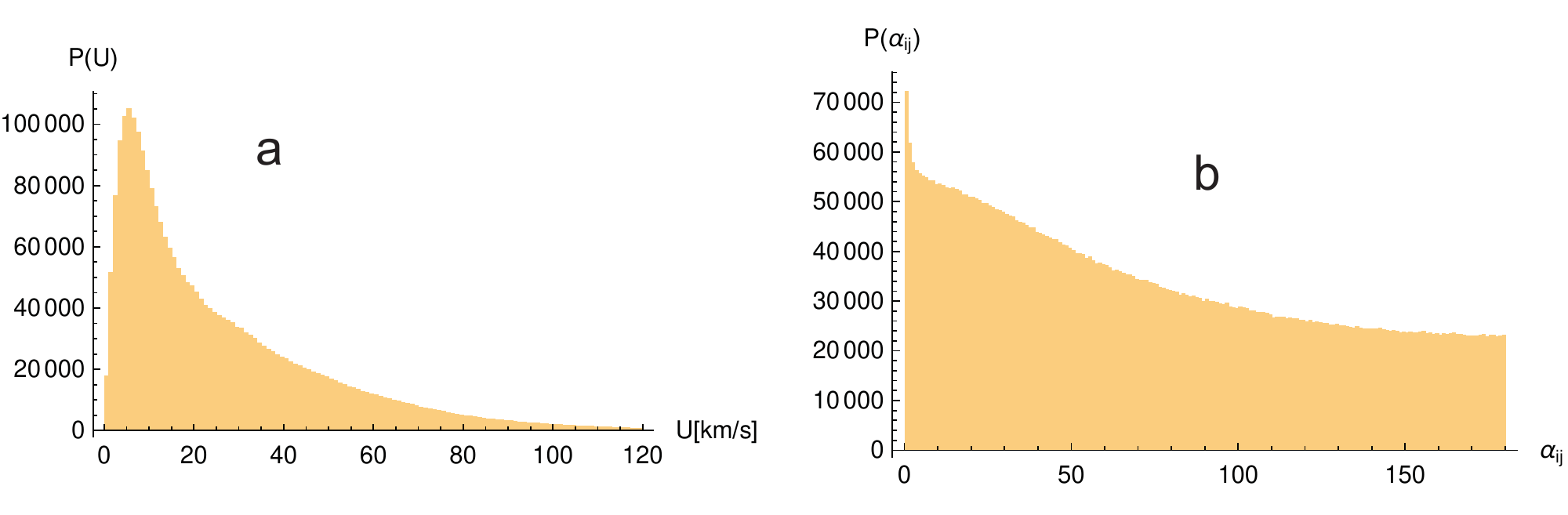}\caption{\textit{Panel a:
}Distribution of the star transverse velocities. \textit{Panels b,c,d:}
Distribution of the corresponding pair angles $\alpha_{ij}$ for $\Delta
_{ij}>0,0.15$ and $1$pc.}%
\label{SAB10}%
\end{figure}

The proper motion of the stars in DR2 is defined by two angular
velocities%
\begin{equation}
\mu_{\alpha\ast}(\equiv\mu_{\alpha}\cos\delta),\qquad\mu_{\delta}
\label{SABE13}%
\end{equation}
in directions of the right ascension and declination in the ICRS. So the
corresponding transverse 2D velocity $\mathbf{U}$ is given as%
\begin{equation}
\mathbf{U=}L\mathbf{(}\mu_{\alpha\ast},\mu_{\delta}),\qquad U=\left\vert
\mathbf{U}\right\vert , \label{SABE14}%
\end{equation}
where $L$ is distance of the star calculated from the parallax (\ref{SABE4b}).
For the pair of stars we can define:%
\begin{equation}
\alpha_{ij}=\arccos\frac{\mathbf{U}_{i}\cdot\mathbf{U}_{j}}{U_{i}U_{j}},\qquad
U_{ij}=\left\vert \mathbf{U}_{i}+\mathbf{U}_{j}\right\vert ,\qquad
v_{ij}=\left\vert \mathbf{U}_{i}-\mathbf{U}_{j}\right\vert , \label{SABE15}%
\end{equation}
where $\alpha_{ij}$\ is angle between both transverse velocities. The
corresponding errors read%
\begin{equation}
\delta\alpha_{ij}\approx\sqrt{2}\frac{\delta u}{u},\qquad\delta U_{ij}=\delta
v_{ij}\approx\sqrt{2}\delta u, \label{SABE15a}%
\end{equation}
where%
\begin{equation}
u=L\mu,\qquad\delta u\approx u\sqrt{\left(  \frac{\delta L}{L}\right)
^{2}+\left(  \frac{\delta\mu}{\mu}\right)  ^{2}}. \label{SABE15b}%
\end{equation}
Note that relative error $\delta v_{ij}/v_{ij}$ can be large, since $v_{ij}$
is small compared with $U_{ij}$ and the errors $\delta v_{ij}$\ and $\delta
U_{ij}$ are the same. In Fig.\ref{SAB10}a we show the distribution of the
velocities $U$ of the stars from the region defined in Tab.\ref{tab1}.
Distribution of the corresponding pair angles $\alpha_{ij}$ is presented in
Fig.\ref{SAB10}b-d for different regions of $\Delta_{ij}$. \ \ \begin{figure}[t]
\centering\includegraphics[width=8cm]{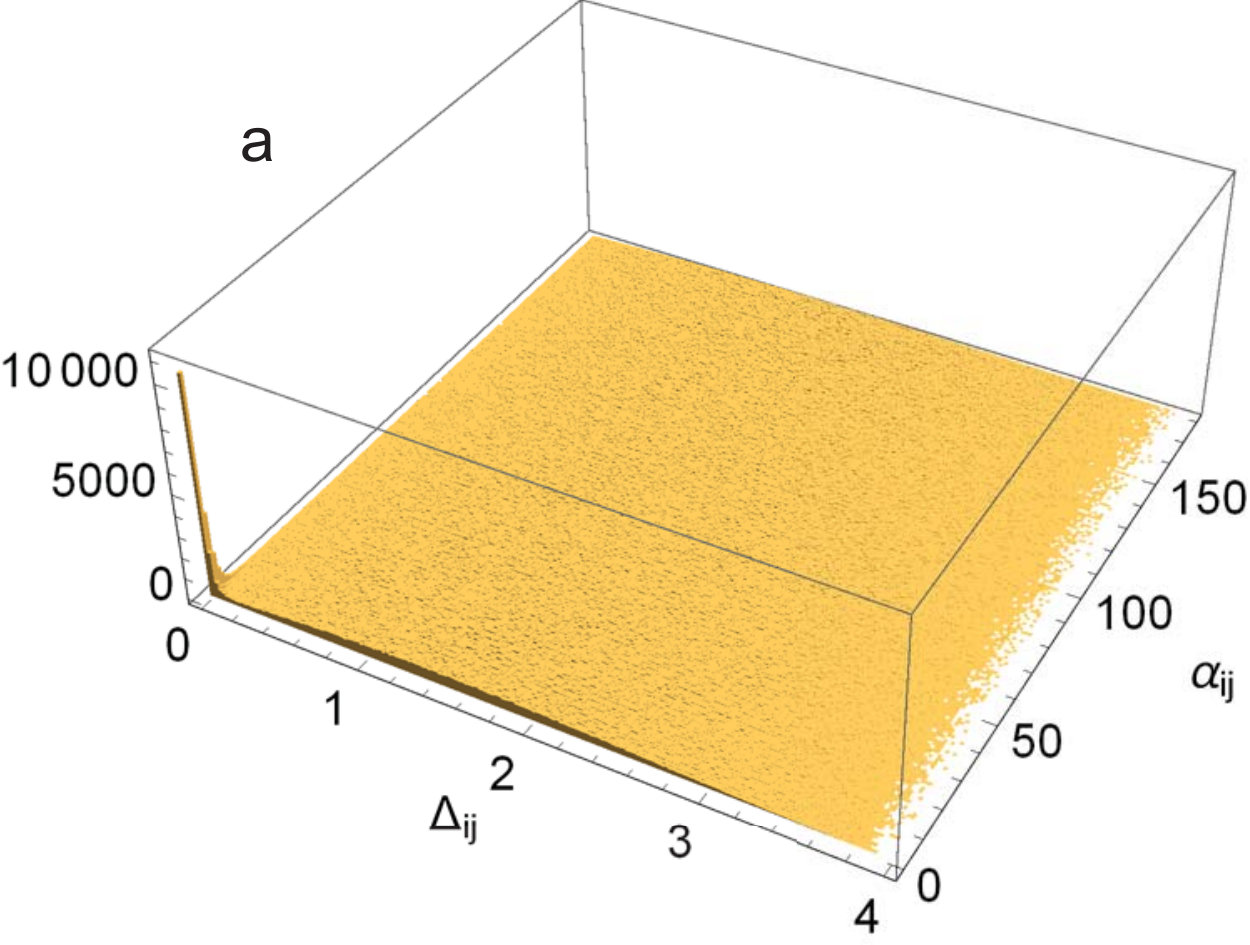}
\includegraphics[width=8cm]{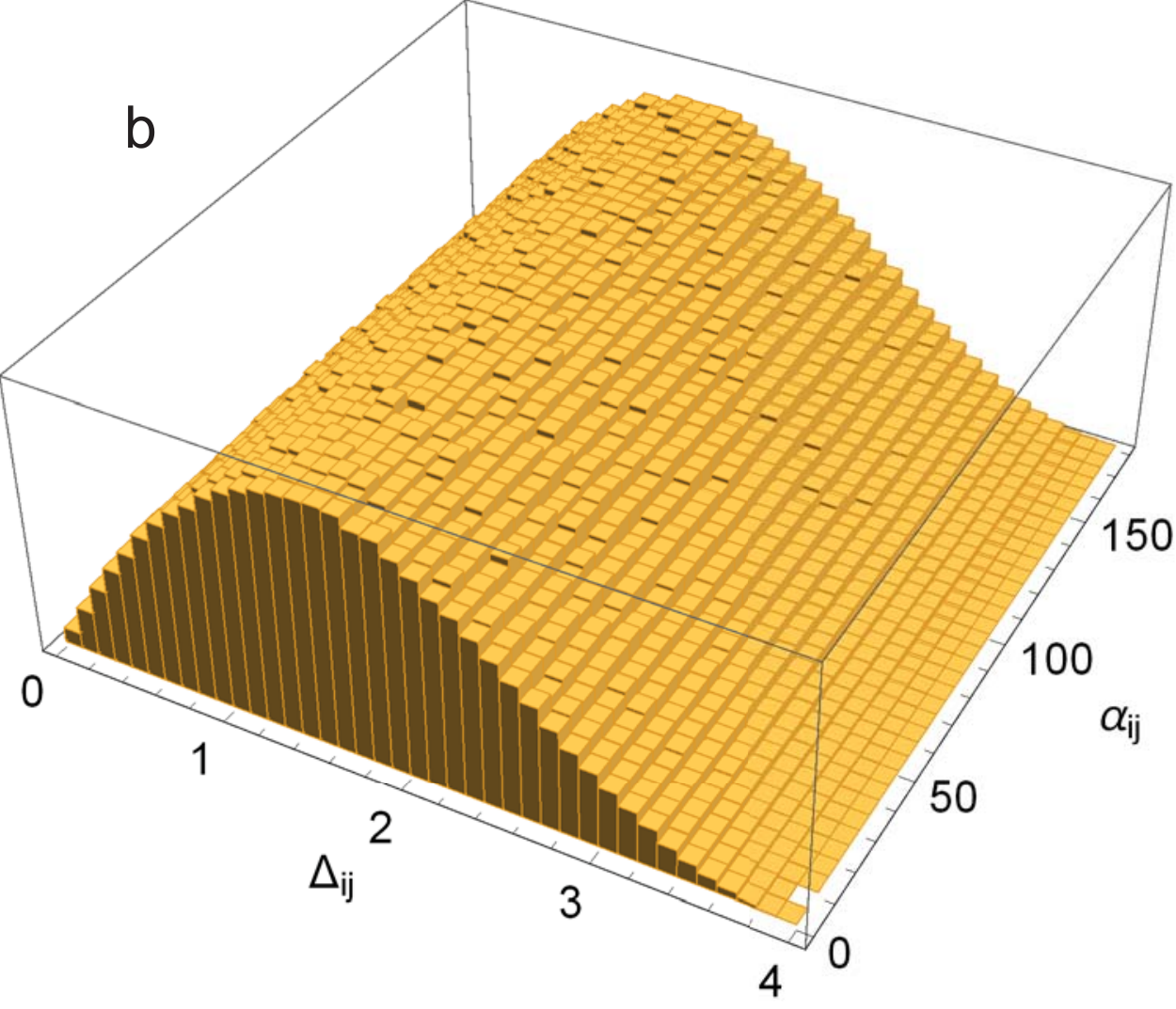}\caption{\textit{Panel a: }Correlation
of the pair transverse separation (unit is 1pc) with angle of corresponding
transverse velocities. \textit{Panel b:} Corresponding MC simulation generated
from uniform distributions of positions and velocity directions.}%
\label{fgr10}%
\end{figure}In Fig.\ref{fgr10}a we have shown the correlation of pair
transverse separations and angles $\alpha_{ij}$. We observe a very narrow peak
in the region of small $\Delta_{ij}$ and $\alpha_{ij}\footnote{Input data
(\ref{SABE13}) are related to the ICRS and $\Delta_{ij}$ is calculated in the
galactic reference frame. Nevertheless the parameters $\Delta_{ij},\alpha
_{ij},U_{ij},v_{ij}$ are invariant under rotation.}$. The peak is connected
with presence of binaries as follows. The transverse velocities of two
gravitationally coupled stars are%
\begin{equation}
\mathbf{U}_{i}=\mathbf{V+v}_{i},\qquad\mathbf{U}_{j}=\mathbf{V+v}_{j},
\label{SABE16}%
\end{equation}
where $\textbf{V}$ is transverse velocity of their center of gravity and
$\textbf{v}_i,\textbf{v}_j$ are transverse projections of instantaneous
orbital velocities, they have always opposite direction. Dominance of very
small $\alpha_{ij}$ means that
\begin{equation}
v_{i}=\left\vert \mathbf{v}_{i}\right\vert \ll\left\vert \mathbf{V}\right\vert
, \label{SABE17}%
\end{equation}
so for binaries in our $\Delta$\ window ($\Delta_{min}$ is given by resolution
of two close sources and $\Delta_{max}=0.1$pc\ by (\ref{SABE9})) we have%
\begin{equation}
U_{ij}\approx2V,\qquad v_{ij}=v_{i}+v_{j}. \label{SABE18}%
\end{equation}
For comparison, we have generated MC plot from uniform distributions of
positions and velocity directions, which is shown in Fig.\ref{fgr10}b.
Obviously, uniform distribution contradicts to Fig.\ref{fgr10}a, which
reflects the presence of binaries (peak at small $\alpha_{ij}$ and \ $%
\Delta_{ij}$). The corresponding $\alpha_{ij}$ peak is also observed in
Fig.\ref{SAB10}b-d on the background of collective motion of stars (dominance
of $\alpha_{ij}<90^\circ$). The peak is suppressed for $\Delta_{ij}\gtrsim1$pc.
Selection of comoving systems is a basis of the methodology applied in the
catalogue JEC.

\begin{figure}[t]
\centering\includegraphics[width=18cm]{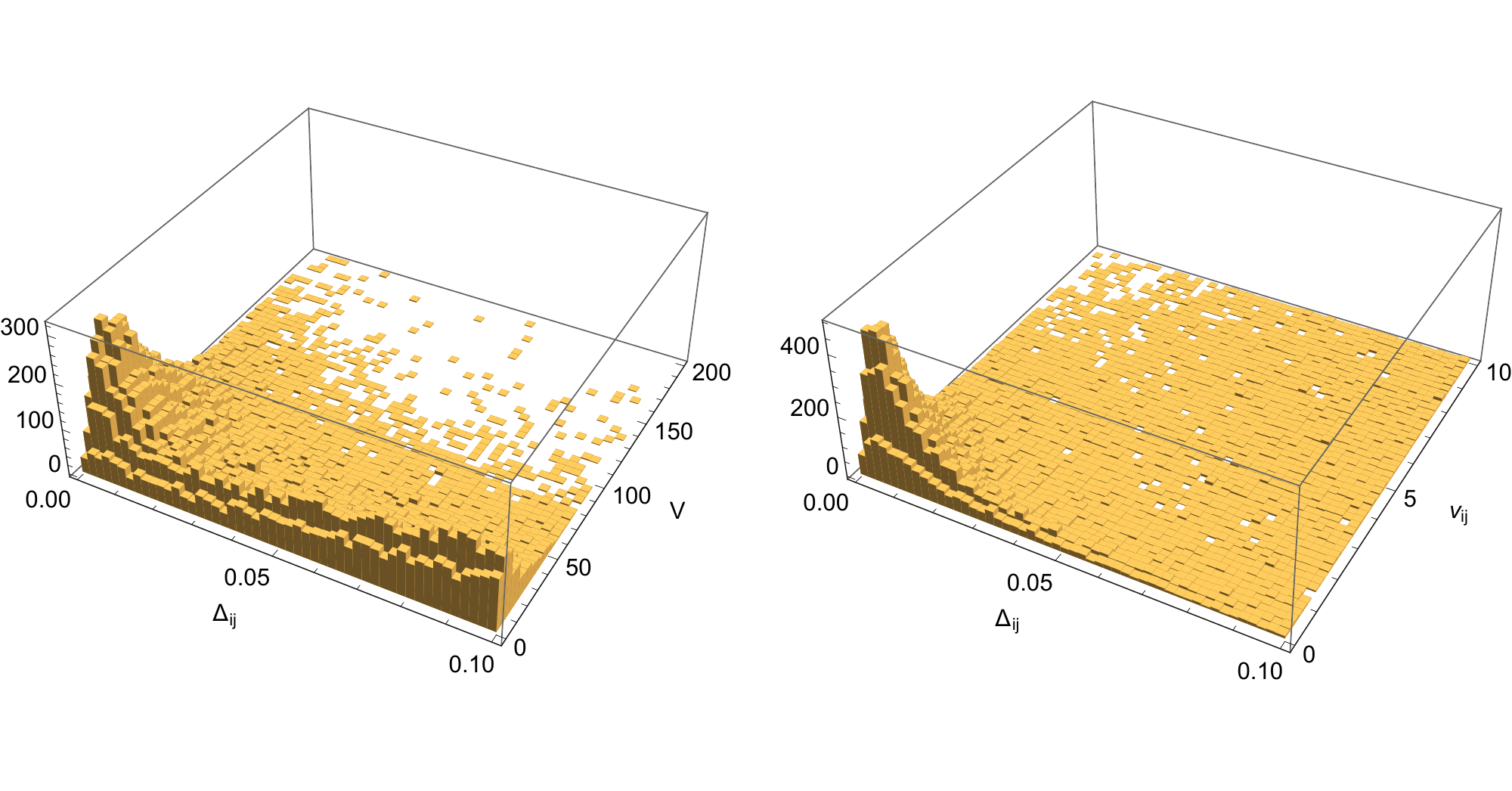}\caption{Correlation of the
pair transverse separation $\Delta_{ij}$ with the transverse velocity $V$ of
the star system and with the transverse velocity $v_{ij}$ of periodic
motion\textit{. }Unit of velocity is km/s.}%
\label{SAB11}%
\end{figure}

In Fig.\ref{SAB11} we show correlations of the velocities $V$ and $v_{ij}$ with
$\Delta_{ij}$ in the region of small separations, where the binaries are
present. With the use of $v_{ij}$ and $\Delta_{ij}$ will try to roughly
estimate the orbital period of the binary star. To simplify the calculation,
we assume the space binary orbits are circular and the star 3D separation is
$a$ (semi-major axis). There are the extreme cases:

A) $M_{1}\approx M_{2}$, then
\begin{equation}
\mathbf{v}_{i}\approx-\mathbf{v}_{j},\qquad v_{ij}\approx v_{i}+v_{j}.
\label{SABE25}%
\end{equation}
Orbital period is%
\begin{equation}
T_{A}\approx\frac{\pi a}{w}, \label{SABE26}%
\end{equation}
where $w$ is the space orbital velocity and $a$ equals to the diameter of the orbit.

B) $M_{1}\gg M_{2}$, then
\begin{equation}
v_{2}\approx0,\qquad v_{ij}\approx v_{1}, \label{SABE27}%
\end{equation}
but the orbital period is different%
\begin{equation}
T_{B}\approx\frac{2\pi a}{w}, \label{SABE28}%
\end{equation}
since the separation $a$ equals to the orbit radius.

\begin{figure}[t]
\centering\includegraphics[width=40mm]{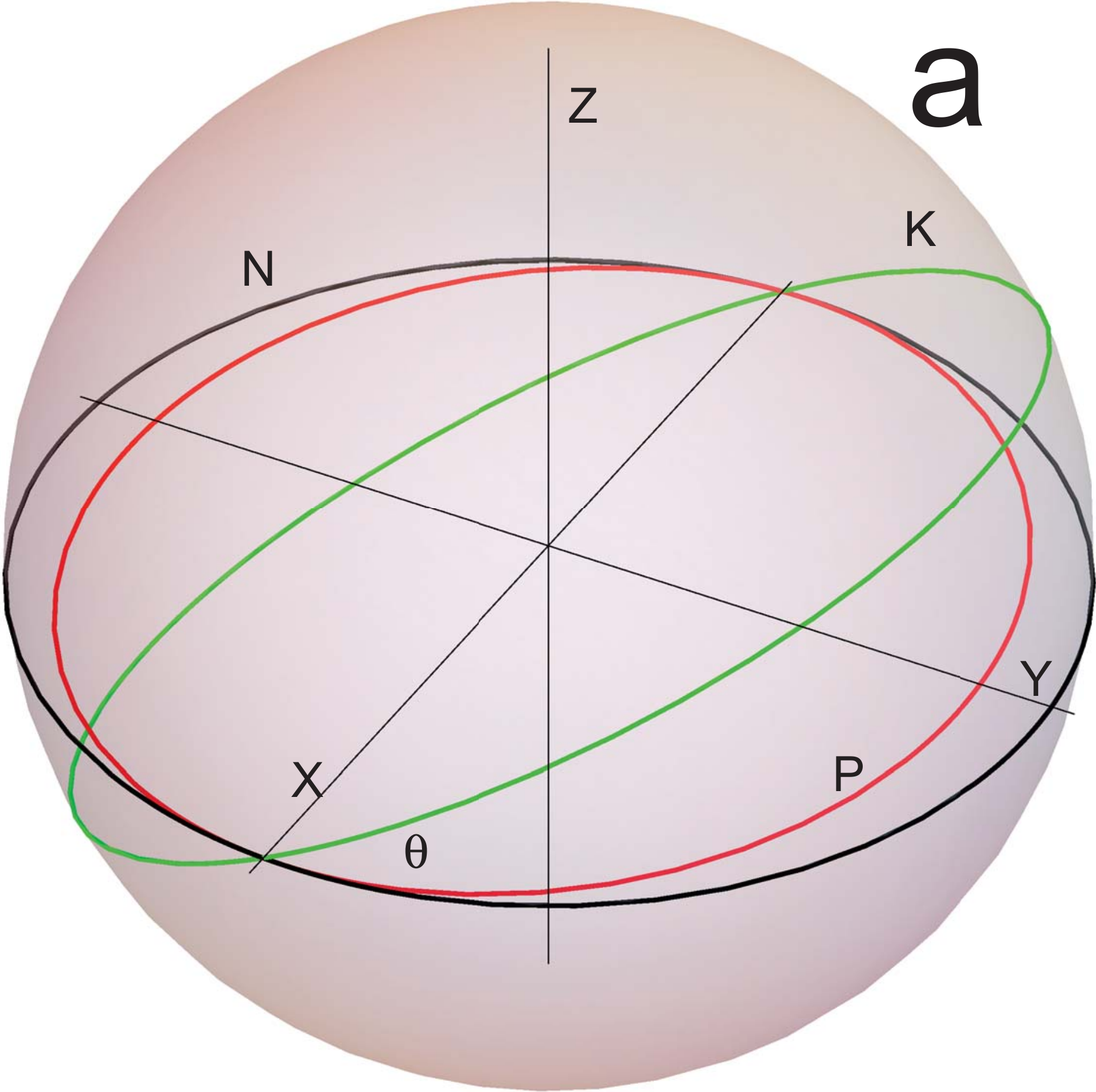}\includegraphics[width=120mm]{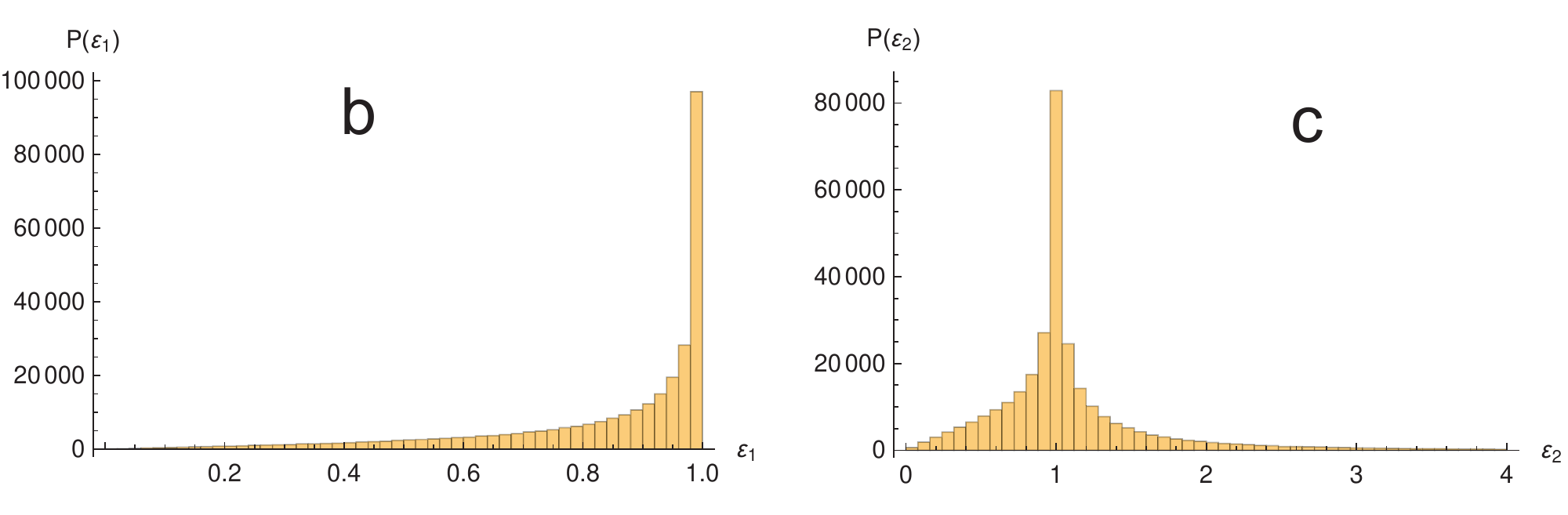}\caption{\textit{Panel
a:} Real orbit (K) and its deformed projection on the sky (P). Real orbit (N)
in the plane $xy.$ \textit{Panels b,c: }Distributions of random parameters
$\varepsilon_{1}$ and $\varepsilon_{2}.$}%
\label{SAB12}%
\end{figure}

At the same time, Kepler's law implies for orbital period $T_{g}:$
\begin{equation}
T_{g}=2\pi\sqrt{\frac{a^{3}}{GM_{tot}}}, \label{SABE10}%
\end{equation}
where $G$ is gravitational constant and$\ M_{tot}=M_{1}+M_{2}$ is mass of the
star system. For units $T_{g}\left[  \text{y}\right]  ,a\left[  \text{pc}%
\right]  $ and $M_{tot}\left[  \text{M}_{\odot}\right]  $ we have%
\begin{equation}
T_{g}=0.937\times10^{8}\sqrt{\frac{a^{3}}{M_{tot}}}. \label{SABE11}%
\end{equation}
This relation also allows us to estimate the period. If the plane of orbit is
perpendicular to the line of sight (axis Z), like the orbit $N$ in
Fig.\ref{SAB12}a, then it is possible to simply substitute:

$v_{ij}/2\rightarrow w,$ in (\ref{SABE26}), $v_{ij}\rightarrow w,$ in
(\ref{SABE28}) and $\Delta_{ij}\rightarrow a$ in (\ref{SABE11}),
(\ref{SABE26}), (\ref{SABE28}).

Then we get:
\begin{equation}
T_{g}=0.937\times10^{8}\sqrt{\frac{\Delta_{ij}^{3}}{M_{tot}}},\qquad
T_{v}=T_{A}=T_{B}=\frac{2\pi\Delta_{ij}}{v_{ij}}. \label{SABE29}%
\end{equation}
The orbit reference frame in figure is similar to the local 3D event frame
defined by the basis (\ref{sa3}) and coordinates (\ref{SABE4a}). But its
origin $\mathbf{L}_{0}$ is defined by the actual position of the centre of
mass of the binary. How to deal with the orbits, whose plane is not
perpendicular to the line of sight like the orbit $K$ in the same figure?
The orbits in the figure are defined as follows:
\[%
\begin{array}
[c]{clll}%
N: & X=r\cos\phi, & Y=r\sin\phi, & Z=0\\
K: & X=r\cos\phi\cos\theta, & Y=r\sin\phi, & Z=r\cos\phi\sin\theta\\
P: & X=r\cos\phi\cos\theta, & Y=r\sin\phi, & Z=0
\end{array}
\]
where $r=a/2$ for the case A, $r=a$ for B and $\phi$ is azimuthal angle in the
plane $XY$. The orbit $K$ inclined at an angle $\theta$ is observed only in
its projection $P$. Corresponding observed separation between the stars is
$\Delta_{ij}:$%
\begin{equation}
a\rightarrow\Delta_{ij},\qquad\sqrt{a^{3}}\rightarrow\sqrt{\Delta_{ij}^{3}%
}=\sqrt{a^{3}}\varepsilon_{1},\qquad\varepsilon_{1}=\left(  \cos^{2}\phi
\cos^{2}\theta+\sin^{2}\phi\right)  ^{3/4}. \label{SABE30}%
\end{equation}
Random angles $\phi,$ $\theta$ \ generate distribution of $\varepsilon_{1} $
shown in Fig.\ref{SAB12}b. The MC distribution demonstrates smearing of the
real separation $a$ due to random $\phi$ and slope $\theta$ of the orbit.
Similarly, the ratio $a/w$ is distorted as%

\begin{equation}
\frac{a}{w}\rightarrow\frac{\Delta_{ij}}{v_{ij}}=\frac{a}{w}\varepsilon
_{2},\qquad\varepsilon_{2}=\frac{\sqrt{\cos^{2}\phi\cos^{2}\theta+\sin^{2}%
\phi}}{\sqrt{\sin^{2}\phi\cos^{2}\theta+\cos^{2}\phi}}. \label{SABE31}%
\end{equation}
Since velocity $\mathbf{w}$ is perpendicular to $\mathbf{r}$, there is
exchange $\cos^{2}\phi\rightleftarrows\sin^{2}\phi$ in denominator.
Corresponding distribution of $\varepsilon_{2}$ is shown in Fig.\ref{SAB12}c.
The mean values are%
\[
\left\langle \varepsilon_{1}\right\rangle =0.791,\qquad\left\langle
\varepsilon_{2}\right\rangle =1.21
\]
and represent a scale of distortion of real orbital periods, if replaced by
relations (\ref{SABE29}). More accurate estimate of the periods in some region
of $\Delta_{ij}$ can be obtained by rescaling of these relations:%
\begin{equation}
\left\langle T_{v}\right\rangle =\frac{2\pi}{\left\langle \varepsilon
_{2}\right\rangle }\left\langle \frac{\Delta_{ij}}{v_{ij}}\right\rangle
,\qquad\left\langle T_{g}\right\rangle =\frac{0.937\times10^{8}}{\left\langle
\varepsilon_{1}\right\rangle \sqrt{M_{tot}}}\left\langle \sqrt{\Delta_{ij}%
^{3}}\right\rangle . \label{SABE32}%
\end{equation}
We have estimated the average periods from the maximum in Fig.\ref{SAB11}. If
we take the sources roughly in the region of half-width of the maximum,
\begin{equation}
\Delta\leq0.015\mathrm{pc},\qquad v_{ij}\leq1.5\mathrm{km/s}, \label{SABE21}%
\end{equation}
then
\begin{equation}
\left\langle T_{v}\right\rangle \approx8.0\times10^{4}\mathrm{y}
\label{SABE22}%
\end{equation}
and one can check that equality $\left\langle T_{v}\right\rangle =\left\langle
T_{g}\right\rangle $ implies estimation $M_{tot}\approx0.8$M$_{\odot}$.

\section{Discussion}

\label{disc}The separation of a pair of stars and the similarity of their
movements can serve as two signatures of the binaries. We can compare them:

\textit{i)} Distributions of the 2D and 3D separations are studied in
Sec.\ref{a2d} and \ref{a3d}. The procedure is simple, the distribution of
separations $P(\Delta)$ (within the defined circles or spheres) is compared
with the corresponding separations $P_{bg}(\Delta)$ generated by uniformly
distributed sources representing background
\begin{equation}
P_{bg}(\Delta)d\Delta=N_{bg}P_{0}(\hat{\xi})d\hat{\xi};\qquad2\rho\hat{\xi
}=\Delta=d_{ij}\ \mathrm{or}\ \Delta_{ij}, \label{sabe23}%
\end{equation}
where $P_{0}$ is given by (\ref{SABE3}) or (\ref{SABE7A})and $N_{bg}$ is the
corresponding number of the background pairs in the data events. Its accurate
calculation is described below, see Eqs. (\ref{sabe42})-(\ref{sabe44}). The
binary distribution reads%
\[
P_{bin}\left(  \Delta\right)  =P\left(  \Delta\right)  -P_{bg}\left(
\Delta\right)
\]
and the probability $\beta\left(  \Delta\right)  $ that the pair is a real
binary is given as
\begin{equation}
\beta\left(  \Delta\right)  =\frac{P_{bin}\left(  \Delta\right)  }{P\left(
\Delta\right)  }=1-\frac{P_{bg}\left(  \Delta\right)  }{P\left(
\Delta\right)  }. \label{sabe35}%
\end{equation}
The function $R\left(  \Delta\right)  $ displayed in Figs. \ref{SAB5}f,
\ref{SAB5}l, \ref{SAB6}c, \ref{SAB7}i and \ref{SAB8}i is another
representation of the probabilistic function $\beta\left(  \Delta\right)  $:
\begin{equation}
R\left(  \Delta\right)  \approx\frac{1}{1-\beta\left(  \Delta\right)  }.
\label{sabe33}%
\end{equation}

\begin{figure}[t]
\centering\includegraphics[width=80mm]{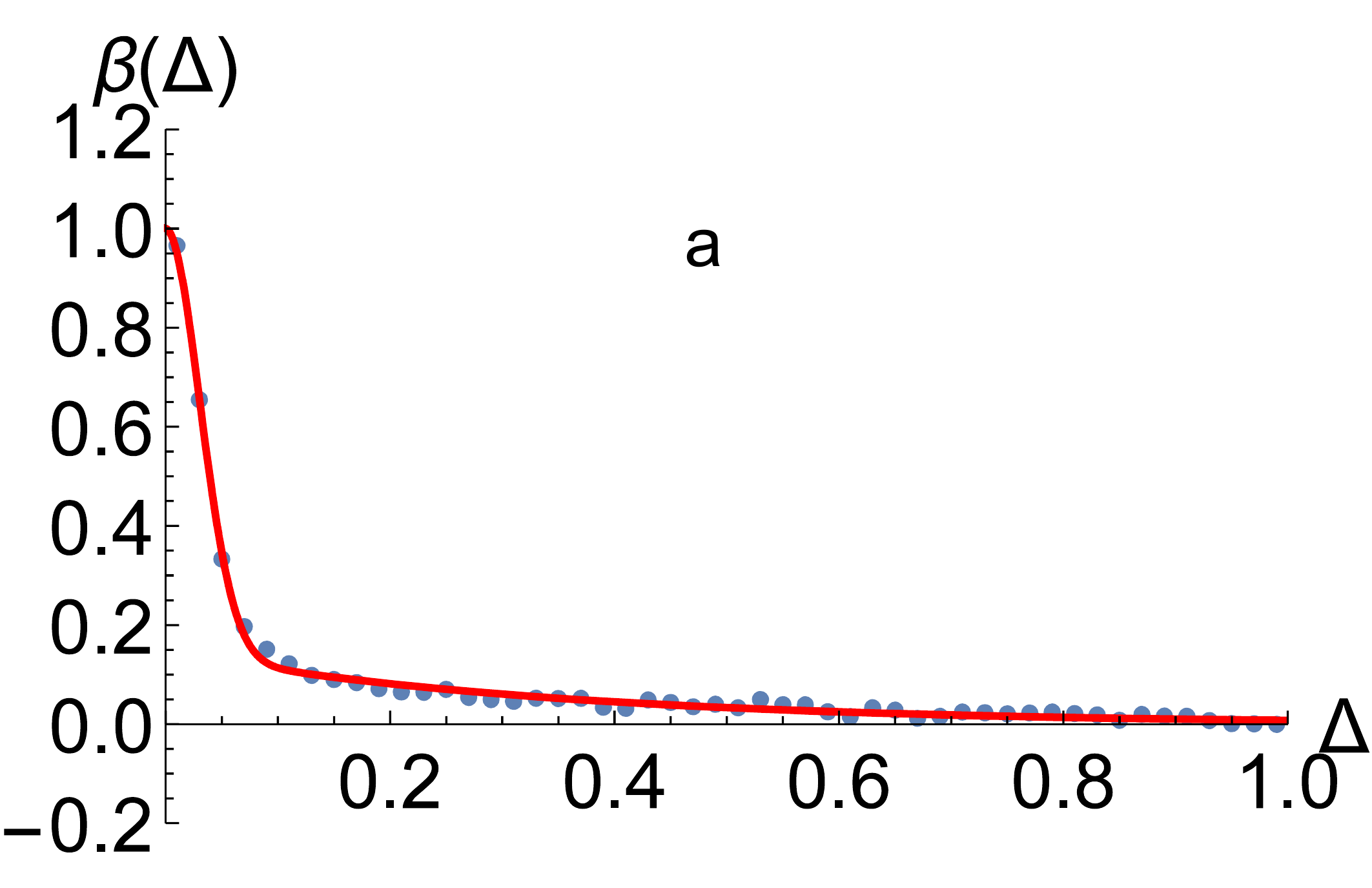}
\includegraphics[width=80mm]{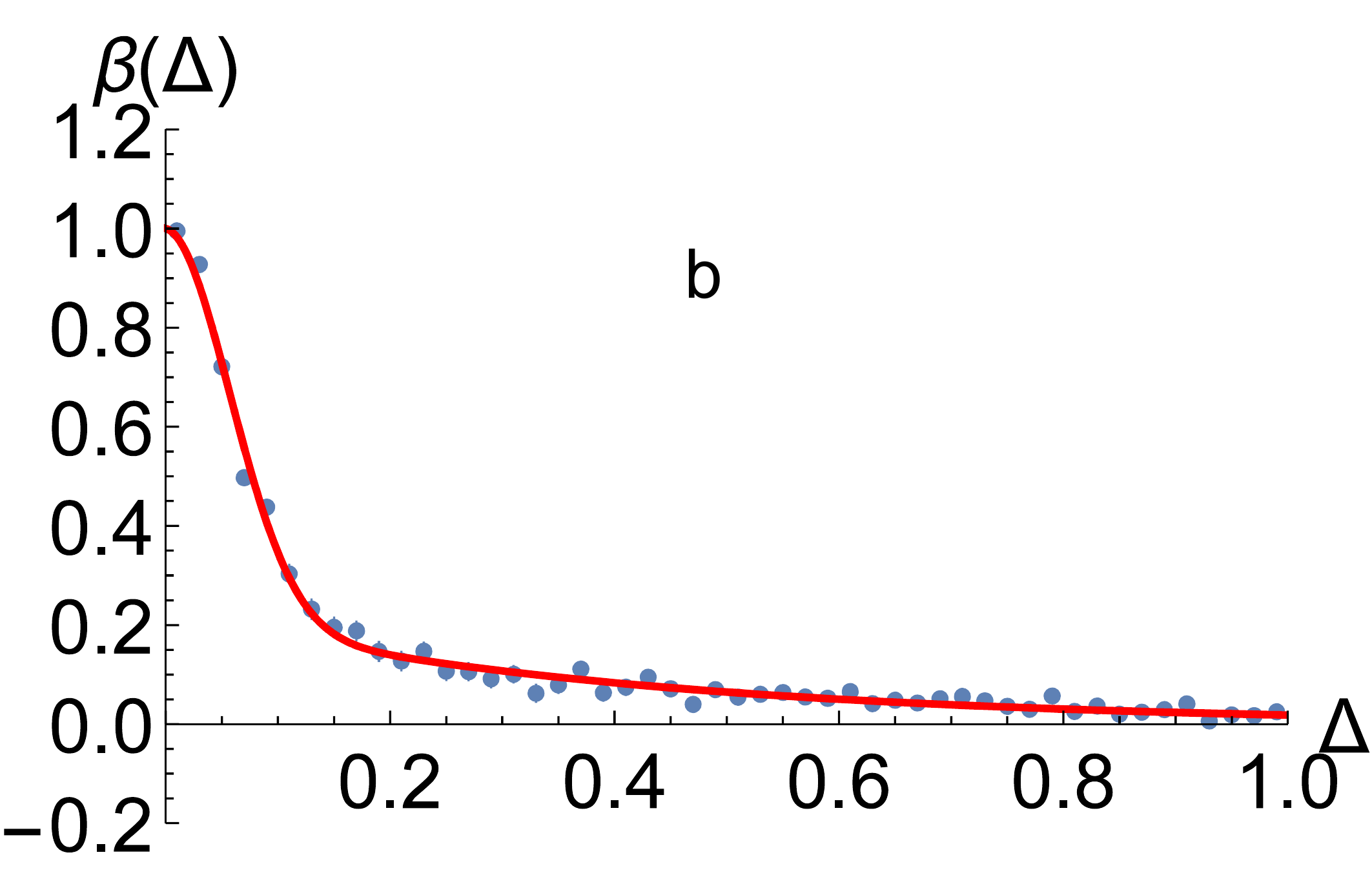}
\includegraphics[width=80mm]{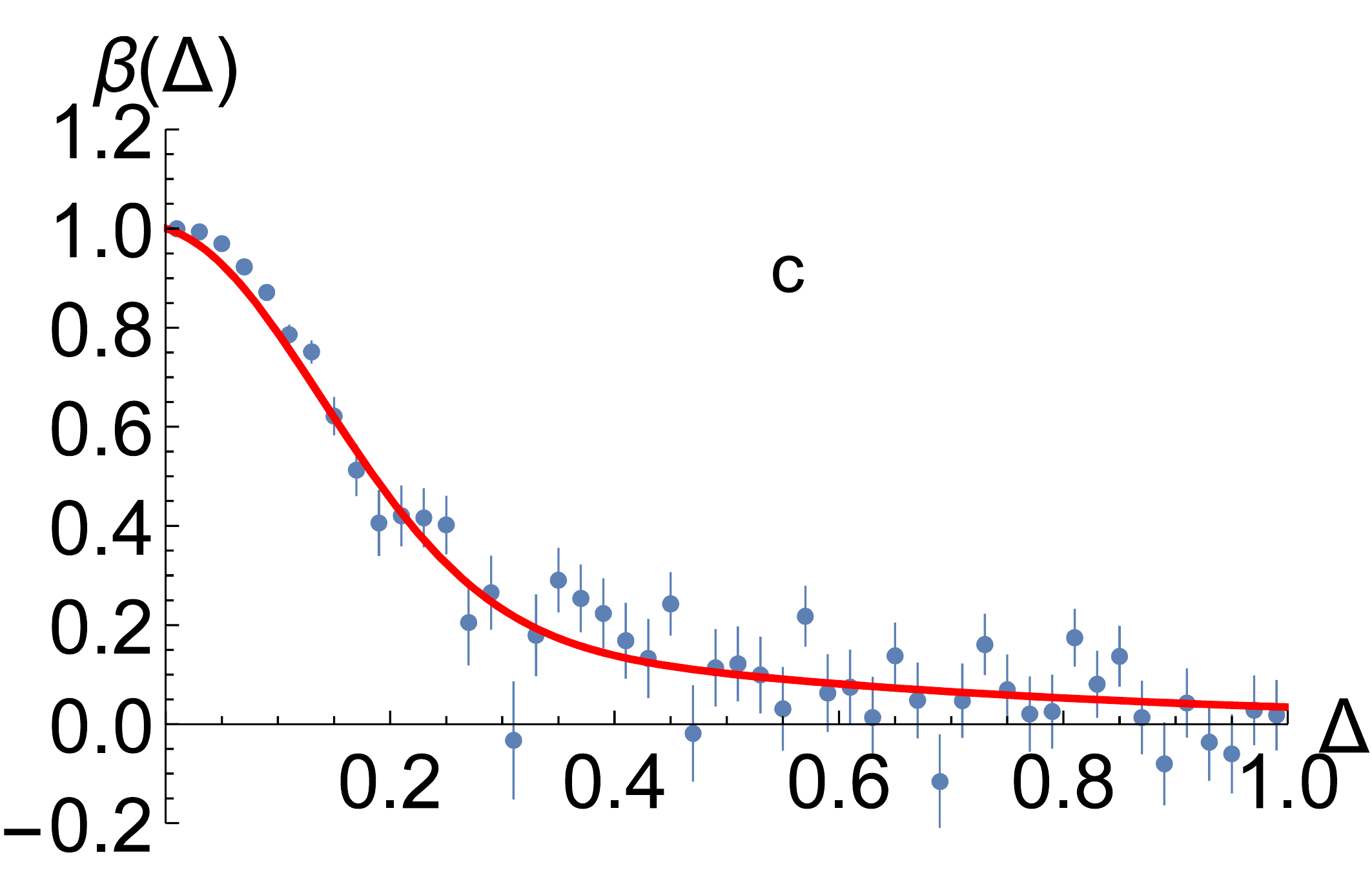}\includegraphics[width=80mm]{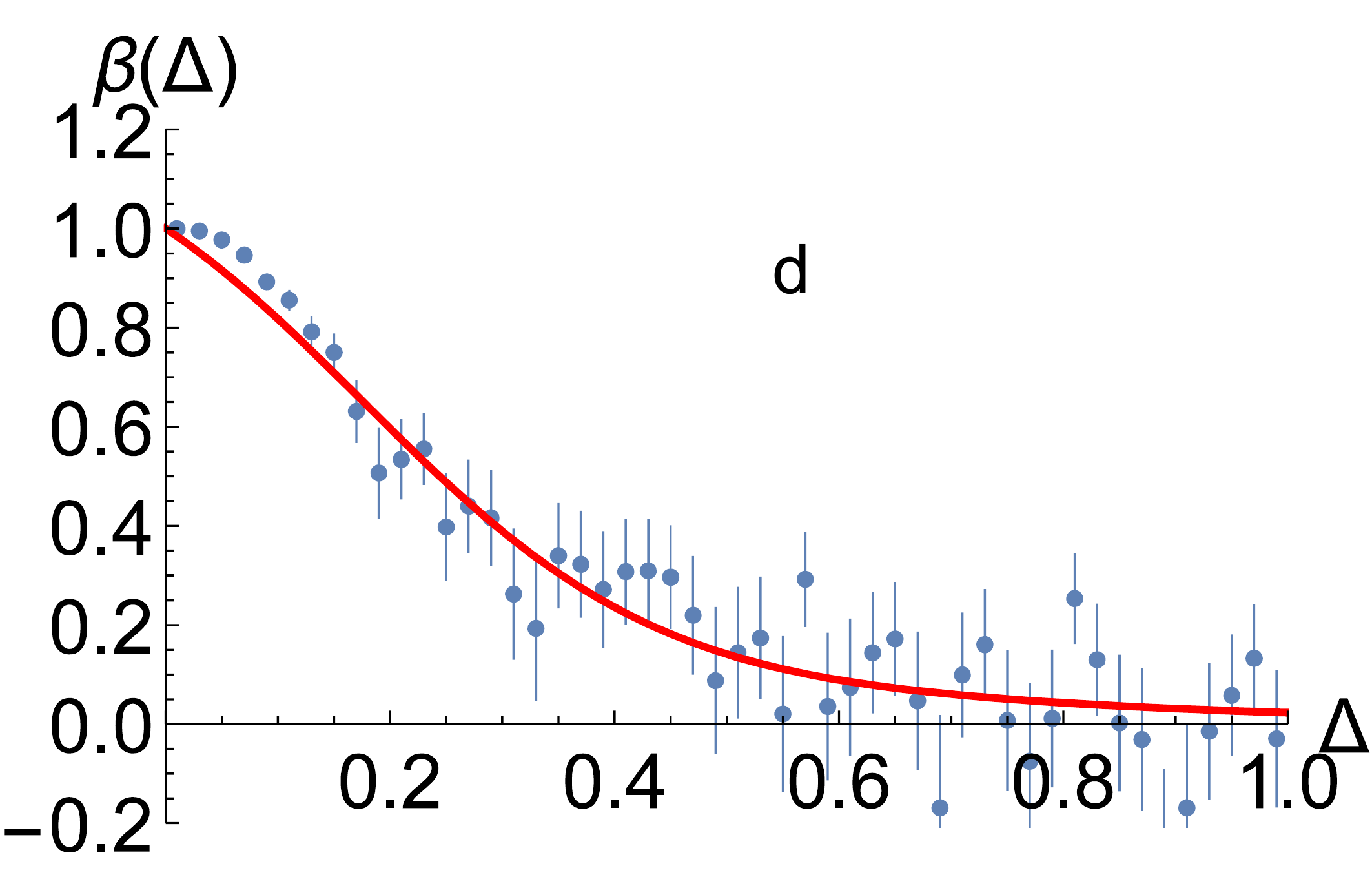}\caption{Probabilistic
function \ $\beta\left(  \Delta\right)  $ for different domains $\Delta
\times\alpha$ and intervals of magnitude $G$. \textit{Panel a:} all pairs and
magnitudes. \textit{Panels \ b,c,d:} united domain  \textbf{AB} for magnitude intervals
$G$ $all,$ $G\leq15,$ $G\leq13$. Blue points with error bars are data, red
curves are fits (\ref{sabe36}).}%
\label{SAB13}%
\end{figure}

Consider the distribution $P\left(  \Delta\right)  $ from Figs.\ref{SAB7}k
and \ref{SAB8}k. The corresponding function $\beta\left(  \Delta\right)  $ is
shown in Fig.\ref{SAB13}a, where the red curve represent fit of the function:%
\begin{equation}
\beta\left(  \Delta\right)  =\omega\exp\left(  -\frac{\Delta^{2}}{2\sigma
_{2}^{2}}\right)  +\left(  1-\omega\right)  \exp\left(  -\frac{\Delta}%
{\sigma_{1}}\right)  . \label{sabe36}%
\end{equation}
The result of the fit is shown in the first row (all pairs) of Tab.\ref{tab2}.
\begin{table}[ptb]
\begin{center}%
\begin{tabular}
[c]{|c|c|c|c|}\hline
& $\omega$ & $\sigma_{1}[pc]$ & $\sigma_{2}[pc]$\\\hline
all pairs & $0.852$ & $0.335$ & $0.0303$\\\hline
all pairs and domain  \textbf{AB} & $0.771$ & $0.396$ & $0.0573$\\\hline
G$\leq$15 and domain  \textbf{AB} & $0.705$ & $0.464$ & $0.143$\\\hline
G$\leq$13 and domain  \textbf{AB} & $0.505$ & $0.326$ & $0.215$\\\hline
\end{tabular}
\end{center}
\caption{Values of the parameters $\omega,\sigma_{1},\sigma_{2}.$}%
\label{tab2}%
\end{table}The longer tail corresponding to the second term describes the
small probability of bound pair at greater separations: $\Delta\gtrsim0.15$pc.
This tail is not visible in panels \textit{i} in Figs. \ref{SAB7} and
\ref{SAB8}. Two exponential terms in $\beta\left(  \Delta\right)  $ may
correspond to two different classes of binaries. The question is to what
extent the excess of wide binaries consists of stable bound systems. Part of
the excess may be an image of widening pairs that were less separated but
weakly bound in the past. The accuracy of the method is based on three conditions:

$\bullet$ precise separation measurement in a suitably selected statistical
set of events that generates $P\left(  \Delta\right)  $

$\bullet$ precise modelling of the background defining $P_{bg}(\Delta)$

$\bullet$ relatively high peak and low background giving probability
$\beta\left(  \Delta\right)  $\ close to 1 in the peak region.

ii) In principle, a similar approach could be applied to comoving pairs.
However, it is obvious the meeting the conditions above is more difficult for
velocities or their differences. The precision of velocity measurement is
lower than separation. The distribution of velocities is far from being simply
uniform, we do not know the exact form of the velocity background.
Fig.\ref{SAB10}b suggests a more complicated collective motion in the
background and a relatively low peak of binaries. That is why we prefer the
primary signature based on the spatial separation, where we work with exactly
defined background and relatively high peaks, like panels \textit{i,k} in
Figs.\ref{SAB7} and \ref{SAB8}. On the other hand, one can expect that a
combination of position and velocity data will suppress background and improve
the selection of binaries in terms of the function $\beta\left(
\Delta\right)  $. A selection based on such a combination is described below.
In Sec.\ref{pmds} we have shown that combination of the space separation with
proper motion allows us to estimate the orbital period.

We denote by $N_{bin}^{D}$ and $N_{bg}^{D}$ the numbers of real and false
(background) binaries in some domain $D$ of separations $\Delta$.
Correspondingly we denote$\ N_{2}^{D}=$ $N_{bin}^{D}+$ $N_{bg}^{D}$, where
$N_{2}^{D}$ is the number of pairs in given dataset. We have%
\begin{equation}
N_{bg}^{D}=\int_{D}P_{bg}\left(  \Delta\right)  d\Delta, \label{sabe42}%
\end{equation}
where $P_{bg}$ is distribution defined by (\ref{SABE7A}) and renormalized in
such a way that%
\begin{equation}
\int_{E}P_{bg}\left(  \Delta\right)  d\Delta=\int_{E}P\left(  \Delta\right)
d\Delta\label{sabe43}%
\end{equation}
where the integration is over the domain $E$ safely outside the peak of
binaries. So we calculate%
\begin{equation}
N_{bg}^{D}=N_{2}^{E}\frac{\int_{D\left(  \Delta\right)  }P_{0}\left(
\Delta\right)  d\Delta}{\int_{E\left(  \Delta\right)  }P_{0}\left(
\Delta\right)  d\Delta}, \label{sabe44}%
\end{equation}
where $P_{0}$\ is defined by (\ref{SABE7A}) with the substitution $\hat{\xi
}=\Delta/2\rho$. The same procedure can be applied for the domains $\ D\left(
\Delta\right)  \times D\left(  \alpha\right)  $. Since we can assume that
$\Delta$ and $\alpha$ are not correlated in the background distribution
\begin{equation}
P_{bg}\left(  \Delta,\alpha\right)  \approx P_{0}\left(  \Delta\right)
P\left(  \alpha\right)  , \label{sabe45}%
\end{equation}
then
\begin{equation}
N_{bg}^{D}=N_{2}^{E}\frac{\int_{D\left(  \Delta\right)  }P_{0}\left(
\Delta\right)  d\Delta}{\int_{E\left(  \Delta\right)  }P_{0}\left(
\Delta\right)  d\Delta}\frac{\int_{D\left(  \alpha\right)  }P\left(
\alpha\right)  d\alpha}{\int_{E\left(  \alpha\right)  }P\left(  \alpha\right)
d\alpha}. \label{sabe46}%
\end{equation}
If we choose $D\left(  \alpha\right)  =E\left(  \alpha\right)  ,$ then the
second ratio is 1. In this way, we can calculate $N_{bg}^{D}$\ without the
knowledge of $P\left(  \alpha\right)  .$ The selected domains are shown,
together with the results $\beta\approx N_{bin}/$ $N_{2}$ in Tab.\ref{tabl4}.
\begin{table}[ptb]
\begin{center}
\centering\includegraphics[width=180mm]{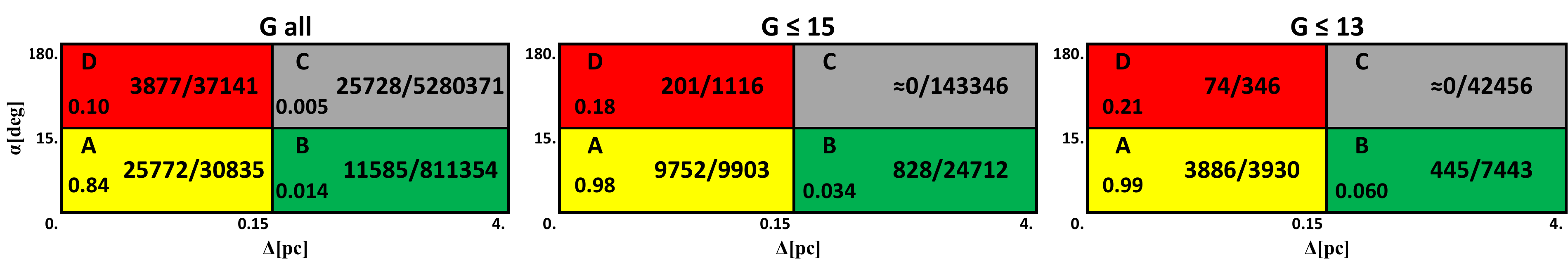}
\end{center}
\caption{The ratio $N_{bin\text{ }}/N_{2\text{ }}$ of the number of binaries
and the total number of pairs in different domains $\Delta\times\alpha$ and
intervals of magnitude $G$.}%
\label{tabl4}%
\end{table}For the calculation, we take the common interval $E\left(
\Delta\right)  \equiv\left\langle 2,4\right\rangle $pc, where $N_{bin}^{E}$
should be zero. In an accordance with Fig.\ref{fgr10}a we observe the highest
rate of binaries in the domain \textbf{A}$\equiv\left\langle
0,0.15\right\rangle $pc$\times\left\langle 0,15\right\rangle $deg. The
binaries are well observable also in the neighbouring domain \textbf{B}%
$\equiv\left\langle 0.15,4\right\rangle $pc$\times\left\langle
0,15\right\rangle $deg. For $G\leq15$ the number of binaries in the domains
\textbf{C} is zero within statistical errors. Presence of binaries in the
domain \textbf{D} requires further analysis.

The probabilistic function $\beta\left(  \Delta\right)  $ corresponding to the
united domain \textbf{AB} is shown in panels \textit{b,c,d} in Fig.\ref{SAB13}
for different intervals of the magnitude $G$. Obviously, the function
(\ref{sabe36}) fitted with the parameters listed in Tab.\ref{tab2} can be
rather approximate. We observe the function $\beta$ is getting wider with
decreasing $G$. This may suggest that brighter and therefore statistically
more massive stars can form stable bound systems even at greater separations.
Obviously, for $\Delta\gtrsim1$pc, the probability $\beta$ is compatible with
zero. This also corresponds to an absence of a binary peak in Fig.\ref{SAB10}%
d, which relates to $\Delta>1$pc. Total numbers of binaries are listed in
Tab.\ref{tab4}. \begin{table}[ptb]
\begin{center}%
\begin{tabular}
[c]{|c|c|c|c|}\hline
& $N_{s}$ & $N_{bin}$ & $N_{bin}/N_{s}$\\\hline
$all\quad G$ & $2734222$ & $66962$ & $0.024$\\\hline
$G\leq15$ & $474689$ & $10781$ & $0.023$\\\hline
$G\leq13$ & $256787$ & $4405$ & $0.017$\\\hline
\end{tabular}
\end{center}
\caption{The number of binaries $N_{bin\text{ }}$ with the total number of
sources $N_{s\text{ }}$ in different intervals of the magnitude $G$.}%
\label{tab4}%
\end{table}The results suggest that in the Gaia DR2 data in the region defined
by Tab.\ref{tab1}, the number of binaries can represent $\approx2\%$ of all
stars in this region.

\section{Catalogue}

\label{cat}In this section, we describe the catalogue of binary candidates,
which we have created from the events of multiplicity $2\leq M\leq15$ defined
by Tab.\ref{tab1}. For the first version of the catalogue we accept only the
candidates from domain \textbf{A} shown in the first panel in Tab.\ref{tabl4}.
So, we do not accept all candidates, but only candidates with a high
probability to be the true binary. The candidates meet the following conditions:

\textbf{1) Projection of separation}

We accept the pairs, which satisfy%

\begin{equation}
\Delta_{cat}\leq0.15\mathrm{pc}. \label{sabe37}%
\end{equation}
In general, the projection of separation $\Delta$ depends on the reference
frame. In the paper, we worked with the local reference frame defined by the
event centre, where projection $\Delta$ into the local plane $XY$ is given by
(\ref{SABE5a}). In the catalogue we do not use local frames. The cut
(\ref{sabe37}) is applied to $\Delta_{cat}$, which is defined as the length of
the arc%
\begin{equation}
\Delta_{ij,cat}=\frac{1}{2}\left(  L_{i}+L_{j}\right)  \arccos\mathbf{n}%
_{i}\mathbf{n}_{j}, \label{sabe40}%
\end{equation}
where $L_{\alpha},\mathbf{n}_{\alpha}$ are defined in section \ref{def}. The
separations $\Delta$ and $\Delta_{cat}$ are not exactly equal, but we have
checked that in our conditions their difference is small, $\left\langle
\delta\Delta\right\rangle \approx0.007$pc. Then the sharp cut on $\Delta
_{cat}$ means only a slightly smeared cut of the distribution of $\Delta$.

\textbf{2) Projection of collinearity}

The pairs must meet the condition
\begin{equation}
\alpha\leq15\deg. \label{sabe41}%
\end{equation}
Both conditions define the domain \textbf{A}\ in the first panel of
Tab.\ref{tabl4}. The panel shows that the average probability of a binary star
is $\beta\approx84\%$. If the stars are brighter, (second and third panel),
then the average $\beta\ $is almost $100\%$.

\textbf{3) Radial separation}

In fact, the radial separation is not explicitly used in our algorithm of
selection of binaries. The reason is a rather low precision of radial
separations, as explained in the comments to Figs. \ref{SAB7} and \ref{SAB8}.
The only constraint is given by the diameter of our events (4pc). Separation
selection is based solely on $\Delta_{ij}$. Additional cuts on inaccurate
radial separation would eliminate many of real binaries and invalidate the
function $\beta$ calculated for $\Delta_{ij}$. We obtained high $\beta$ even
without a cut on radial separation.

Further, it is evident that spherical events fill the space only partially
($\approx52\%$). In this way, half the stars is lost for analysis. We also
lose binaries between two adjacent events when each star falls into another.
In order to recover these losses, we work with the \ modified coverage:

\textit{i)} The event spheres are replaced by cubes of edge 4pc with no gaps
between them. In each cube, we search for the pairs meeting the conditions
(\ref{sabe37}) and (\ref{sabe41}).

\textit{ii)} The procedure is repeated with the same cubes centred in the
corners of the former cubes and the search results are merged.

The catalogue is represented by a matrix that is defined as follows. Each line
represents one star and there are the following data in the columns:

1--2: \ Group ID and Group size $\left(  n\geq2\right)  $ to match stars with
the group they belong to\footnote{In the current version of our catalogue we
omit candidates of systems n%
$>$
2, so only binaries (n = 2) are written.}.

3--96: \ Copy of the original entry for the star from Gaia-DR2 archive
\footnote{\url{http://cdn.gea.esac.esa.int/Gaia/gdr2/gaia_source/csv/}},
according to the documentation
\footnote{\url{http://gea.esac.esa.int/archive/documentation/GDR2/Gaia_archive/chap_datamodel/sec_dm_main_tables/ssec_dm_gaia_source.html}}%
.

97--98: \ Minimum and maximum angular separation of the star from other stars
in the group [$as$].

99--100: \ Minimum and maximum projected physical separation of the star from
other stars in the group [pc].

The summary data from our catalogue of binary candidates (I), along with the
data extracted from the catalogue JEC - \cite{esteban} (II) is shown in
Tab.\ref{tab5}. The number of candidates $N_{2}$ in this table correspond to
$N_{2}$ in the domains \textbf{A} in Tab.\ref{tabl4} after increasing with the
repeated covering. \begin{table}[ptb]
\begin{center}%
\begin{tabular}
[c]{|c|c|c|}\hline
Catalogue & $N_{2}\quad\left(  N_{bin}\right)  $ & $G$\\\hline
\multicolumn{1}{|l|}{(I): $\ \ \ \ \ $domain  \textbf{A}} & $80560\quad\left(
67670\right)  $ & $all$\\\hline
\multicolumn{1}{|l|}{} & $22674\quad\left(  22201\right)  $ & $\leq15$\\\hline
& $9082\quad\left(  8991\right)  $ & $\leq13$\\\hline
\multicolumn{1}{|l|}{(II): \ total \ $N_{bin}=3055$} & $\left(  301\right)  $
& $\leq13$\\\hline
\end{tabular}
\end{center}
\caption{Summary table of binaries from the catalogues (I) and (II), see text.
$N_{2}$ is the number of binary candidates, $N_{bin}$ is the real expected
number of binaries. The data in second column are related to the full cube
region $\left(  400\mathrm{pc}\right)  ^{3}$ (Tab.\ref{tab1}).}%
\label{tab5}%
\end{table}The comparison of the two catalogues shows the following:

a) We can only compare sources of magnitude $G\leq13,$ because (II) does not
contain less bright stars. Of the total number $3055$ binaries in (II), only
$301$ lie in the (I) cube $\left(  400\mathrm{pc}\right)  ^{3}$. Increasing
the edge of this cube by the factor $2.15$ would increase volume $10$ times
with the number of binaries comparable to the total (II).

b) The number $301$(II) could be compared with the corresponding number of
candidates $C_{2}=9082$(I). However, in the JEC catalogue, apart from the cut
$G\leq13$ many other restrictions and selections are made. Definition of the
binary is not the same in both catalogues. In our opinion, this is the reason
for the difference between (I) and (II).

c) (I) and (II) have $108$ common binary candidates.

d) 86(II) candidates are absent in (I) since the separation $\Delta$ exceeds
0.15pc. These candidates do not contradict to our general criteria, but the
corresponding probability $\beta\left(  \Delta\right)  $ can be lower as shown
in Fig.\ref{SAB13}.

e) 54(II) candidates are absent in (I) since their spatial separation exceeds
4pc (event diameter). Such candidates may not contradict to our criteria,
however $\beta\left(  \Delta\right)  $ due to a great background can be
extremely low.

f) 18(II) candidates are absent in (I) since these candidates are located in a
dense area generating the high multiplicity events, which exceed our currently
set limit.

g) The last 35(II) candidates are absent in (I), mainly due to the fact that
even after the second coverage some couples (II) remain separated in two
neighbouring event cubes.

The total number of the binary candidates of all Gaia magnitudes in the
catalogue (I) is $80560$, which corresponds to the expected real number of
binaries $67670$. The full current catalogue (I) in the csv form is available
on the website \url{https://www.fzu.cz/~piska/Catalogue/}. We plan to further
develop and optimize our catalogue methodology.

\section{Summary and conclusion}

\label{sum}We have proposed a general statistical method for analysis of
finite 2D and 3D patterns. In the present study, the method has been applied
to the analysis of binary star systems in different regions of the Gaia
catalogue DR2.

Results on 2D statistical analysis were compared with our former results
obtained from the previous catalogue DR1. The new results give in the
distribution of angular separations more clear evidence of binaries.
Independent signature follows from the characteristic functions $\Theta
_{n}(M)$, which clearly indicate a tendency to clustering. However, the most
important results are obtained from the 3D analysis introduced in the present
paper. We have analyzed about $5\times10^{5}$ of events inside the cube of
edge 400pc centred at the origin of the galactic reference frame. In
distributions of pair separations, we observe the sharp peaks at small
separations corresponding to binaries, which are more striking for brighter
sources, $G\leq15$.

The important result of the analysis is probabilistic function $\beta\left(
\Delta\right)  $, which depends on the separation $\Delta$ of a pair of stars
and indicates the probability that the pair constitutes a bound system. The
function suggests that brighter, more massive binary stars have on average a
greater separation. With increasing separation the function falls rapidly. We
obtained the ratio binaries/singles $\approx2\%$.

Further, we had shown that\ a combined analysis of 3D separations with the
proper motion of the pairs of sources gives a clear picture of the binaries
with two components of the motion: parallel and orbital. The analysis allowed
us to estimate the average orbital period and mass of the binary star system
in the chosen statistical ensemble.

The highest probability of the binary is observed at smallest separations
$\Delta$ and angles $\alpha$ between proper motions. From the corresponding
domain $\Delta\times\alpha\equiv\left\langle 0,0.15\right\rangle $%
pc$\times\left\langle 0,15\right\rangle $deg, we have created the catalogue
involving $80560$ binary candidates, which represents $67670$ of the true
binaries. \begin{figure}[h]
\centering\includegraphics[width=60mm]{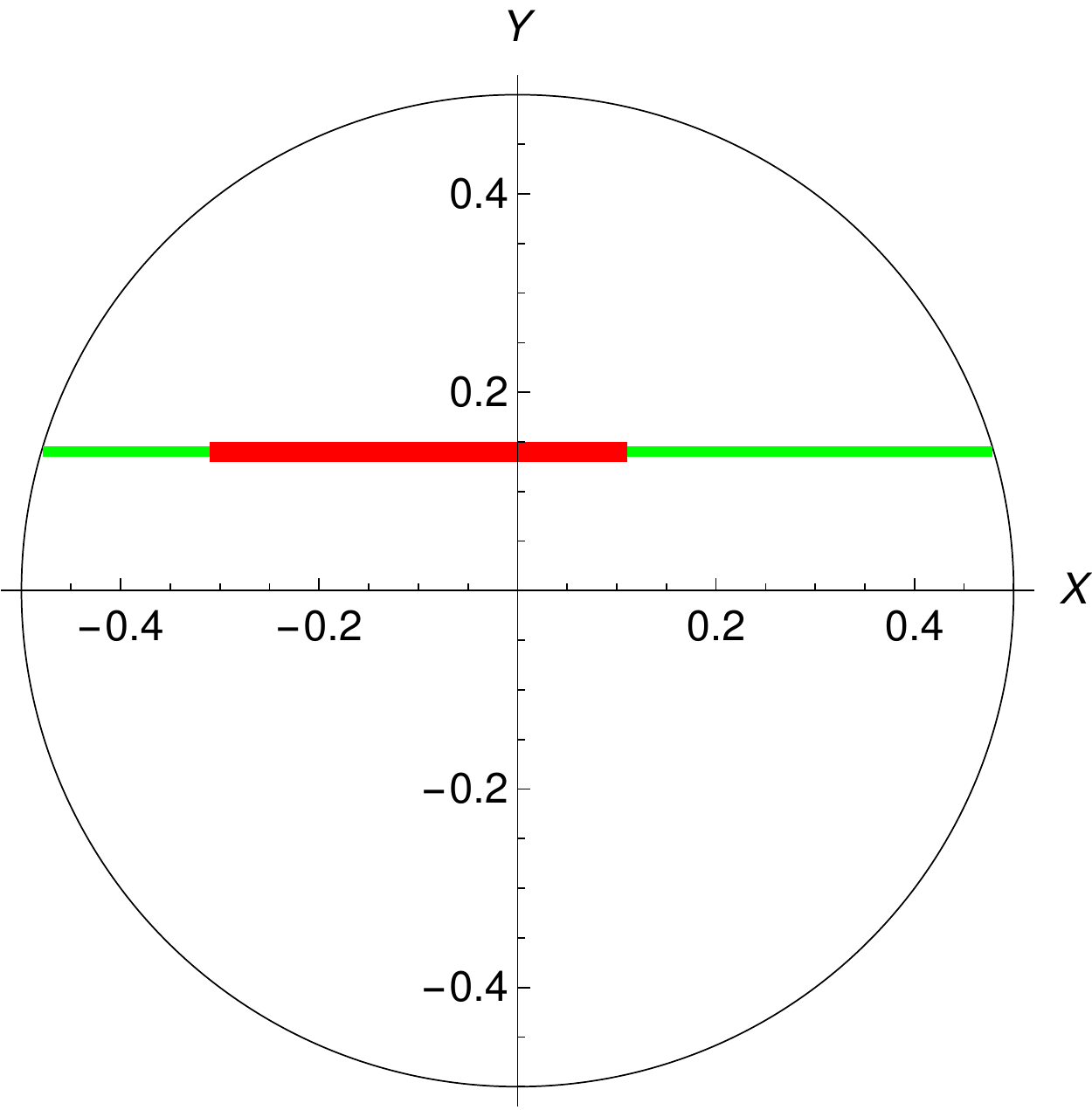} \ \caption{Random segment
$l$ (red) on the chord $L(y)$ (green).}%
\label{ape}%
\end{figure}

\appendix\textbf{\ }

\section{\textbf{Proof of relations }(\ref{SABE3}), (\ref{SABE4}),
(\ref{SABE7}), (\ref{SABE7A}) and (\ref{SABE8})}

\label{appe}

\textit{i) Relation (\ref{SABE3})}

First we consider two random points on a segment $L$. The probability that the
points are separated by interval $l$ reads:%
\begin{equation}
p(l)\sim L-l. \label{ap1}%
\end{equation}
Further, we suppose a circle of diameter $2R=1$ with a chords $L(y)$ involving
random segment $l$ (Fig.\ref{ape}). We have%
\begin{equation}
L(y)=2\sqrt{\left(  1/2\right)  ^{2}-y^{2}}=\sqrt{1-\left(  2y\right)  ^{2}.}
\label{ap2}%
\end{equation}
The probability of interval $l$ reads
\begin{align}
P_{\parallel}(l)  &  \sim\int_{0}^{y_{\max}}\left(  L(y)-l\right)
dy\sim\left(  \arccos l-l\sqrt{1-l^{2}}\right)  ;\label{ap3}\\
y_{\max}  &  =\frac{1}{2}\sqrt{1-l^{2}},\quad0<l<1.\nonumber
\end{align}
This distribution is generated by random pairs on the chords parallel to axis
$x$. For arbitrary random pairs separated by $l$ inside the circle, we
integrate distributions (\ref{ap3}) over all directions in 2D and replace
\begin{equation}
dl\rightarrow d\left(  \pi l^{2}\right)  \sim ldl, \label{ap5}%
\end{equation}
which gives distribution
\begin{equation}
P_{\parallel}(l)dl\rightarrow P(l)dl\sim P_{\parallel}(l)ldl,\quad0<l<1.
\label{ap4}%
\end{equation}
Relation (\ref{SABE3}) is its normalized form.

\textit{ii) Relation (\ref{SABE4})}

Distribution (\ref{ap4}) can be modified%
\begin{equation}
P(l)dl\sim P_{\parallel}(l)ldl\sim P_{\parallel}\left(  \sqrt{l_{x}^{2}%
+l_{y}^{2}}\right)  dl_{x}dl_{y}. \label{ap7}%
\end{equation}
Calculation of integral%
\begin{equation}
P(l_{x})\sim\int_{0}^{l_{y{\max}}}P_{\parallel}\left(  \sqrt{l_{x}^{2}%
+l_{y}^{2}}\right)  dl_{y},\quad l_{y{\max}}=\sqrt{1-l_{x}^{2}} \label{ap8}%
\end{equation}
with the use of \ \cite{wolfram} and after replacement $l_{x}\rightarrow
\hat{\xi}$ and normalization gives relation (\ref{SABE4}).

\textit{iii) Relation (\ref{SABE7})}

Now instead of circle $2R=1,$ we consider the sphere of the same radius. The
procedure is a modification of the case\textit{\ i)}. Now instead of integral
(\ref{ap3}) we get%
\begin{align}
P_{\parallel}(l)  &  \sim\int_{0}^{y_{\max}}y\left(  L(y)-l\right)
dy\sim\left(  \frac{1}{3}-\frac{l}{2}+\frac{l^{3}}{6}\right)  ;\label{ap9}\\
y_{\max}  &  =\frac{1}{2}\sqrt{1-l^{2}},\quad0<l<1,\nonumber
\end{align}
where $y$ means radius of a cylinder of parallel chords. The additional $y$ in
the integral means that we integrate chords on surfaces of cylinders of
different radii. Then instead of (\ref{ap5}) we use%
\begin{equation}
dl\rightarrow d\left(  \frac{4}{3}\pi l^{3}\right)  \sim l^{2}dl, \label{ap10}%
\end{equation}
since the integration of chords is over all directions in 3D. Resulting
distribution reads%

\begin{equation}
P(l)dl\sim P_{\parallel}(l)l^{2}dl\sim l^{2}\left(  \frac{1}{3}-\frac{l}%
{2}+\frac{l^{3}}{6}\right)  , \label{ap11}%
\end{equation}
which after normalization gives relation (\ref{SABE7}).

\textit{iv) Relation (\ref{SABE8})}

Probability $\mathbf{P}\left(  l\right)  $ of random segments $l=\sqrt
{l_{x}^{2}+l_{y}^{2}+l_{z}^{2}}$ inside the sphere can be expressed as%
\begin{equation}
\mathbf{P}\left(  \sqrt{l_{x}^{2}+l_{y}^{2}+l_{z}^{2}}\right)  dl_{x}%
dl_{y}dl_{z}\sim\mathbf{P}\left(  l\right)  l^{2}dl\sim P(l)dl, \label{ap14}%
\end{equation}
which together with (\ref{ap11}) gives
\begin{equation}
\mathbf{P}\left(  \sqrt{l_{x}^{2}+l_{y}^{2}+l_{z}^{2}}\right)  =\mathbf{P}%
\left(  l\right)  \sim\left(  \frac{1}{3}-\frac{l}{2}+\frac{l^{3}}{6}\right)
. \label{ap15}%
\end{equation}
The probability that segment $l$ has projection $l_{x}$ is given as%
\begin{align}
P(l_{x})  &  \sim\int_{0}^{l_{z{\max}}}\int_{0}^{l_{y{\max}}}\mathbf{P}\left(
\sqrt{l_{x}^{2}+l_{y}^{2}+l_{z}^{2}}\right)  dl_{y}dl_{z}\label{ap12}\\
&  \sim\int_{0}^{t_{\max}}\mathbf{P}\left(  \sqrt{l_{x}^{2}+t^{2}}\right)
tdt;\quad t_{\max}=\sqrt{1-l_{x}^{2}}. \label{ap13}%
\end{align}
The last integral (equally for $l_{y},l_{z}$) can be after inserting from
(\ref{ap15}) easily calculated, and after normalization gives relation
(\ref{SABE8}).

\textit{v) Relation (\ref{SABE7A})}

In a similar way, the probability that segment $l$ has projection
$\Delta=\sqrt{l_{x}^{2}+l_{y}^{2}}$ is given as%
\begin{equation}
P(\Delta)\sim\int_{0}^{l_{z{\max}}}\mathbf{P}\left(  \sqrt{\Delta^{2}%
+l_{z}^{2}}\right)  \Delta dl_{z};\quad l_{z{\max}}=\sqrt{1-\Delta^{2}},
\label{ap16}%
\end{equation}
which after inserting from (\ref{ap15}) and integration with the use of
\ \cite{wolfram} implies relation (\ref{SABE7A}).

\acknowledgments
This work has made use of data from the European Space Agency (ESA) mission
\textit{Gaia} (\url{https://www.cosmos.esa.int/gaia}), processed by the
\textit{Gaia} Data Processing and Analysis Consortium (DPAC,
\url{https://www.cosmos.esa.int/web/gaia/dpac/consortium}). Funding for the
DPAC has been provided by national institutions, in particular the
institutions participating in the \textit{Gaia} Multilateral Agreement. The
work was supported by the project LTT17018 of the MEYS (Czech Republic).
Further, we are grateful to J.Grygar for deep interest and many valuable
comments and J.Palou\v{s} and O.Teryaev for\ very useful discussions and
inspiring comments.


\begin{thebibliography}{99999999999999999999999999999999}                                                                 %


\bibitem[Arenou(2018)]{arenou2}Arenou, F., Luri, X., Babusiaux, C., et al.
2018, A\&A, 616, A17


\bibitem[Cabalero(2009)]{cabalero}Caballero, J. A. 2009, A\&A, 507, 251

\bibitem[Close et al.(1990)]{close}Close, L. M., Richer, H. B., \& Crabtree,
D. R. 1990, AJ, 100, 1968

\bibitem[Gaia Collaboration(2016b)]{gaia}Gaia Collaboration (Prusti T. et al.)
2016, A\&A, 595, A1


\bibitem[Gaia Collaboration(2016a)]{gaia1}Gaia Collaboration (Brown, A. G. A.
et al.) 2016, A\&A, 595, A2


\bibitem[Gaia Collaboration(2018)]{gaia2}Gaia Collaboration (Brown, A. G. A.,
et al.) 2018, A\&A, 616, A1


\bibitem[Jim{\'e}nez-Esteban et al.(2019)]{esteban}Jim{\'e}nez-Esteban, F.~M.,
Solano, E. \& Rodrigo, C. 2019, AJ, 157, 78

\bibitem[Mathematica(2018)]{wolfram}Wolfram Research, Inc., Mathematica,
Version 11.3, Champaign, IL (2018)

\bibitem[Oelkers(2017)]{oelkers}Oelkers, R. J., Stassun, K. G., \& Dhital, S.
2017, AJ, 153,259

\bibitem[Oh(2017)]{semyeong}Oh, S., Price-Whelan, A. M., Hogg, D. W., Morton,
T. D., \& Spergel, D. N. 2017, AJ, 153, 257

\bibitem[Zavada\&P\'{\i}\v{s}ka(2018)]{AApzkp}Zavada P. \& P\'{\i}\v{s}ka K.,
2018, A\&A, 614, A137


\bibitem[Ziegler(2018)]{ziegler}Ziegler C. et al. 2018, AJ, 156,259
\end{thebibliography}
\end{document}